\documentclass[prx,twocolumn,superscriptaddress,notitlepage,nofootinbib]{revtex4-2}
\usepackage[utf8]{inputenc}
\usepackage[T1]{fontenc}
\usepackage[english]{babel}
\usepackage[dvipsnames]{xcolor}
\usepackage{lmodern,fancyhdr}
\usepackage{graphicx,float}
\usepackage{enumitem,amsmath,amssymb,amsthm,bm,array,mathdots,mathrsfs,dsfont,bbold}
\usepackage[colorlinks]{hyperref}
\hypersetup{colorlinks,linkcolor=red,citecolor=blue,urlcolor=blue,final}
\usepackage{verbatim}
\usepackage[normalem]{ulem}
\newcommand{\bra}[1]{\langle {#1} | }
\newcommand{\ket}[1]{| {#1} \rangle }
\newcommand{\pol}[1]{\boldsymbol{\varepsilon}_{#1}}
\newcommand{\polK}{\boldsymbol{\varepsilon}_{\mathbf{k}}}
\newcommand{\sumkappa}{\int d\mathbf{k}\sum_{\boldsymbol{\varepsilon}_\mathbf{k}}}
\newcommand{\im}{\text{i}}

\newcommand{\be}{\begin{equation}}
\newcommand{\ee}{\end{equation}}
\newcommand{\eqnref}[1]{Eq.~\eqref{#1}}
\newcommand{\figref}[1]{Fig.~\ref{#1}}
\newcommand{\secref}[1]{Sec.~\ref{#1}}

\newcommand{\keps}{\mathbf{k},\boldsymbol{\varepsilon}_\mathbf{k}}

\begin{document}
\title{Suppressing Recoil Heating in Levitated Optomechanics using Squeezed Light}

\author{C.~Gonzalez-Ballestero}
\affiliation{Institute for Quantum Optics and Quantum Information of the Austrian Academy of Sciences, A-6020 Innsbruck, Austria.}
\affiliation{Institute for Theoretical Physics, University of Innsbruck, A-6020 Innsbruck, Austria.}
\email{carlos.gonzalez-ballestero@uibk.ac.at}

\author{J.~A.~Zielińska}
\affiliation{Photonics Laboratory, ETH Zürich, CH-8093 Zürich, Switzerland}

\author{M. Rossi}
\affiliation{Photonics Laboratory, ETH Zürich, CH-8093 Zürich, Switzerland}

\author{A.~Militaru}
\affiliation{Photonics Laboratory, ETH Zürich, CH-8093 Zürich, Switzerland}
\author{M. Frimmer}
\affiliation{Photonics Laboratory, ETH Zürich, CH-8093 Zürich, Switzerland}

\author{L. Novotny}
\affiliation{Photonics Laboratory, ETH Zürich, CH-8093 Zürich, Switzerland}

\author{P.~Maurer}
\affiliation{Institute for Quantum Optics and Quantum Information of the Austrian Academy of Sciences, A-6020 Innsbruck, Austria.}
\affiliation{Institute for Theoretical Physics, University of Innsbruck, A-6020 Innsbruck, Austria.}

\author{O.~Romero-Isart}
\affiliation{Institute for Quantum Optics and Quantum Information of the Austrian Academy of Sciences, A-6020 Innsbruck, Austria.}
\affiliation{Institute for Theoretical Physics, University of Innsbruck, A-6020 Innsbruck, Austria.}

\begin{abstract}
We theoretically show that laser recoil heating in free-space levitated optomechanics can be arbitrarily suppressed by shining  squeezed light onto an optically trapped nanoparticle. The presence of squeezing modifies the quantum electrodynamical light-matter interaction in a way that enables us to control the amount of information that the scattered light carries about a given mechanical degree of freedom. Moreover, we analyze the trade-off between measurement imprecision and back-action noise and show that optical detection beyond the standard quantum limit can be achieved. We predict that, with state-of-the-art squeezed light sources, laser recoil heating can be reduced by at least $60\%$ by squeezing a single Gaussian mode with an appropriate incidence direction, and by $98\%$ by squeezing a properly mode-matched mode. Our results, which are valid both for motional and librational degrees of freedom, will lead to improved feedback cooling  schemes as well as  boost the coherence time of optically levitated nanoparticles in the quantum regime.
\end{abstract}

\maketitle

\section{Introduction}

Recent experiments have achieved active feedback ground-state cooling of a center-of-mass degree of freedom of an optically levitated dielectric nanosphere in free space, that is, in the absence of an optical cavity~\cite{MagriniNature2021,TebbenjohannsNature2021,KambaOptExp2022,AikawaArxiv2022}. This achievement requires (i) to efficiently measure the light scattered by the nanoparticle in order to extract all the information it carries about the center-of-mass position, and (ii) that the motional noise is dominated by the measurement back-action, that is, laser recoil heating~\cite{JainPRL2016}. After ground-state cooling, the coherence time of the quantum mechanical degree of freedom of an optically levitated nanoparticle is still limited by (ii), even though the collection of the scattered photons (i) is not needed anymore. To achieve coherence times beyond what laser recoil heating allows, one needs to either (a) switch off the laser light, a strategy followed in the context of macroscopic quantum physics~\cite{RomeroIsartPRL2011,RomeroIsartPRA2011b,BatemanNatCom2014,neumeier_FastQuantum_2022,RodaLlordesarXiv2023} and hybrid particle control with electric and magnetic forces~\cite{GonzalezBallesteroScience2021,RomeroIsartPRL2012,MillenPRL2015,FonsecaPRL2016,SlezakNJP2018,ConanglaNanoLett2020,DaniaPRR2021,HoferArXiv2022,GutierrezLatorrearXiv2022}, or (b) suppress the amount of information about the nanoparticle position that is carried by the scattered light. Option (b) is especially useful for free-space experiments as it would enable to suppress laser-induced noise while keeping the laser, and thus the trapping potential, unchanged. Conversely, the complementary ability to \emph{increase} the amount of information carried by the scattered light would improve the sensitivity of optical measurements of the particle position, thus allowing to cool its motion to lower temperatures.

\begin{figure}[t!]
	\centering
	\includegraphics[width=\linewidth]{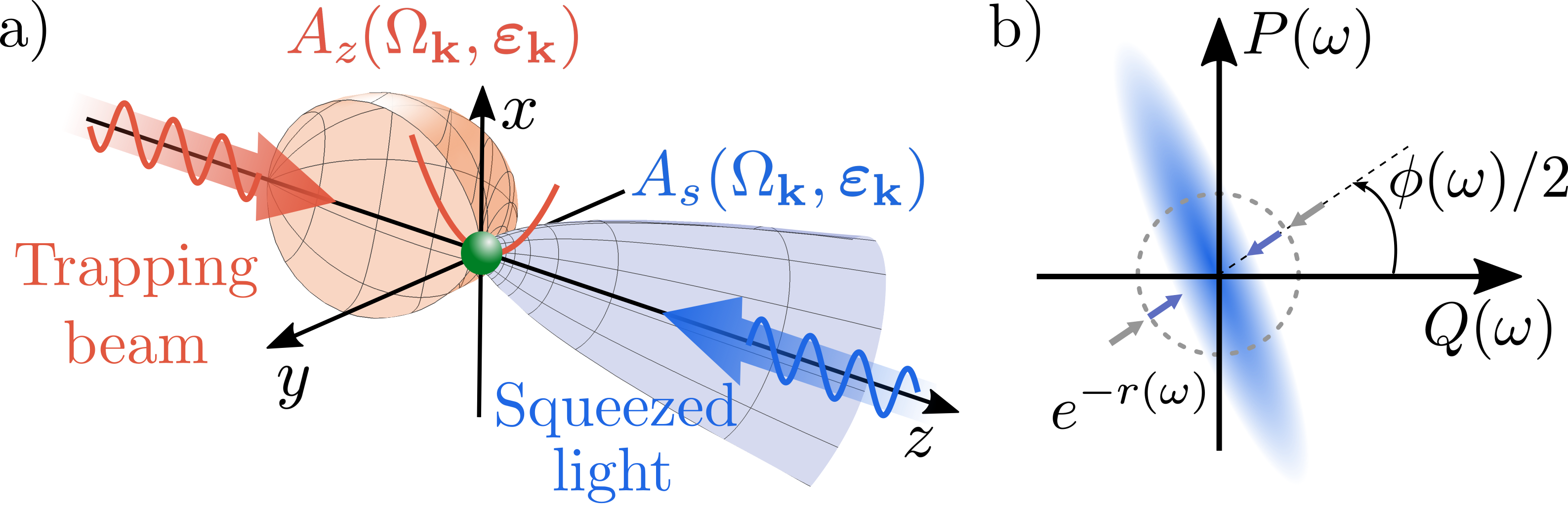}
	\vspace{-0.5cm}
	\caption{a) We consider a sub-wavelength particle (green) trapped by a focused laser beam (red arrow) and interacting with an electromagnetic mode in a squeezed vacuum state (blue arrow). The functions $A_{s}(\Omega_\mathbf{k},\pol{\mathbf{k}})$ and $A_{z}(\Omega_\mathbf{k},\pol{\mathbf{k}})$ describe, respectively, the angular and polarization distribution of the squeezed beam and of the photons scattered inelastically by the motion along $z$ in the absence of squeezing.
	b) Wigner function of the squeezed light mode at frequency $\omega$, see Appendix~\ref{AppWigner} for details.
	}\label{Figure1}
	\vspace{-0.4cm}
\end{figure}

In this work, we propose a method to suppress recoil heating or to enhance optical detection sensitivity for an optically trapped nanoparticle using squeezed light (\figref{Figure1}). First, we theoretically show that by squeezing a particular electromagnetic mode that interacts with the optically trapped nanoparticle (blue beam in \figref{Figure1}), the quantum electrodynamical light-matter interaction can be modified to reduce the information carried by the scattered light about a given mechanical degree of freedom. Consequently, laser recoil heating is reduced by an amount proportional to the degree of squeezing. We predict that the recoil heating can be fully suppressed using a properly designed squeezed beam and, moreover, that it can be reduced by at least $60\%$ in current experiments. 
Second, we model the optical detection of particle motion and discuss the trade-off between measurement imprecision and back-action noise as a function of the squeezing parameters. We show that the sensitivity of optical measurements can lie beyond the standard quantum limit (SQL). In analogy to related ideas in the context of single-mode cavity optomechanics~\cite{AsjadPRA2016} and electromechanics~\cite{ClarkNature2017}, this fact opens the door to using  squeezed light to reach lower temperatures via feedback cooling of an optically levitated nanoparticle. Our work allows to extend early proposals to use squeezed light to enhance detection sensitivity in cavity- and 1D optomechanics~\cite{CavesPRD1981,Unruh1983,BondurantPRD1984,Jaekel1990,PacePRA1993} to continuum, cavity-less, free-space optomechanics. 

This article is organised as follows. In Sec.~\ref{secHamiltonian} we derive the light-matter Hamiltonian in the presence of squeezed light. Then, in Sec.~\ref{secRecoilInfo}, we compute the motional recoil heating rates and the squeezing-induced modification of the angular dependence of the scattered photons (information radiation patterns~\cite{TebbenjohannsPRA2019, MaurerQED2}). In Sec.~\ref{secDETECTION} we derive a theoretical model for photodetection of the particle motion, namely we derive input-output relations for the light modes and compute the minimum mechanical signal detectable in the output light. We then extend all the above results from center-of-mass motion to the librational degrees of freedom of optically trapped nanorotors in Sec.~\ref{secROTORA}. Finally, our conclusions are presented in Sec.~\ref{SecConclusion}.

\section{Hamiltonian in the presence of squeezing}\label{secHamiltonian}

In this section we introduce the light-matter Hamiltonian and the core quantities characterizing its modification due to squeezing. First, we summarize the derivation of the linearized optomechanical Hamiltonian for a sub-wavelength dielectric particle in free space in the absence of squeezing (Sec.~\ref{SecHbare}). Then, we derive its expression in the presence of squeezed light generated by spontaneous parametric down conversion, by introducing a suitably chosen basis of collective electromagnetic modes (Sec.~\ref{SecHsqueezing}).

\subsection{Light-Matter Hamiltonian in the absence of squeezing}\label{SecHbare}

We consider a sub-wavelength dielectric particle of mass $m$ and dielectric permittivity $\epsilon$ trapped at the focus of a laser beam of frequency $\omega_0$ propagating along the positive $z-$axis and polarized along the $x-$axis (\figref{Figure1}a).
We choose the origin of coordinates at the equilibrium position of the particle. The fundamental light-matter interaction Hamiltonian reads 
\begin{equation}
    \hat{H}=\frac{\hat{\mathbf{p}}^2}{2m} + \hat{H}_{\rm EM} + \hat{H}_{\rm int},
\end{equation}
and contains the kinetic energy of the particle with center-of-mass momentum $\hat{\mathbf{p}}$, the Hamiltonian of the electromagnetic field $\hat{H}_{\rm EM}$, and the interaction between the field and the center-of-mass motion $\hat{H}_{\rm int}$. In the small-particle limit the Hamiltonian of the field can be well approximated by its diagonal plane-wave expansion~\cite{MaurerQED1},
\begin{equation}\label{HemBARE}
    \hat{H}_{\rm EM} \approx \hbar\sumkappa \omega(k) \hat{a}^\dagger(\keps)\hat{a}(\keps),
\end{equation}
with $\omega(k) \equiv ck\equiv c\vert\mathbf{k}\vert$ (where $c$ is the speed of light in vacuum), $\mathbf{k}\in\mathds{R}^3$ being the wave vector of each plane-wave mode, and $\boldsymbol{\varepsilon}_\mathbf{k}\perp\mathbf{k}$ denoting each of the two orthogonal polarization vectors for each $\mathbf{k}$. The electromagnetic ladder operators fulfil bosonic commutation relations, $[\hat{a}(\keps),\hat{a}^\dagger(\mathbf{k}',\pol{\mathbf{k}'})]=\delta(\mathbf{k}-\mathbf{k}')\delta_{\polK,\pol{\mathbf{k}'}}$.
Similarly, in the small-particle limit the light-matter coupling  can be described by the interaction Hamiltonian of a polarizable point-dipole in an electric field, $\hat{H}_{\rm int}=-(\alpha/2)\hat{\mathbf{E}}^2(\hat{\mathbf{r}})$~\cite{RomeroIsartNJP2010,ChangPNAS2010,RomeroIsartPRA2011,GonzalezBallesteroPRA2019,MaurerQED2}, with $\hat{\mathbf{r}}$ being the center-of-mass position operator, $\alpha\equiv3\epsilon_0 V(\epsilon-1)/(\epsilon+2)$ the particle polarizability, $V$ its volume, and $\hat{\mathbf{E}}(\mathbf{r})$ the electric field operator.

In the presence of a highly populated laser with frequency $\omega_0$, the above nonlinear Hamiltonian can be linearized~\cite{RomeroIsartNJP2010,ChangPNAS2010,RomeroIsartPRA2011,GonzalezBallesteroPRA2019,MaurerQED2}. As detailed in Appendix~\ref{AppA}, this results in the following Hamiltonian in the absence of squeezing, 
\begin{multline}
\label{H0}
    \frac{\hat{H}^{(0)}}{\hbar}=\sum_\mu\Omega_\mu\hat{b}^\dagger_\mu\hat{b}_\mu +\sumkappa \Delta(k)\hat{a}^\dagger(\keps)\hat{a}(\keps) \\
    +\sum_\mu\hat{q}_\mu \int d\mathbf{k}\sum_{\boldsymbol{\epsilon}_\mathbf{k}}\left[G_\mu(\mathbf{k},\boldsymbol{\epsilon}_\mathbf{k})\hat{a}^\dagger(\mathbf{k},\boldsymbol{\epsilon}_\mathbf{k}) +\text{H.c.}\right],
\end{multline}
with $\Delta(k)\equiv \omega(k)-\omega_0$. The first term describes the Hamiltonian of the center-of-mass motion along the three Cartesian axes in terms of bosonic ladder operators, $[\hat{b}_\mu,\hat{b}_{\mu'}]= \delta_{\mu\mu'}$, and the mechanical frequencies $\Omega_\mu$ originating from the optical restoring force. The second term describes the electromagnetic field Hamiltonian in the rotating frame. The third term in \eqnref{H0} describes the light-matter interaction in terms of the dimensionless displacement operator $\hat{q}_\mu\equiv\hat{r}_\mu/r_{0\mu} =\hat{b}_\mu+\hat{b}_\mu^\dagger$, with $r_{0\mu}\equiv [\hbar/(2m\Omega_\mu)]^{1/2}$ the zero-point motion, and of the linearized coupling rate $G_\mu(\mathbf{k},\boldsymbol{\epsilon}_\mathbf{k})$ in the small-particle limit. 

The general form of the coupling rates is well-known~\cite{TebbenjohannsPRA2019,AndreiArxiv2022,MaurerQED2} and given in Appendix~\ref{AppA}. However, under the approximations performed in this work, all relevant quantities depend only on the coupling rate evaluated at the laser frequency, given by~\cite{supplementary,MagriniNature2021,GonzalezBallesteroPRA2019,AndreiArxiv2022,MaurerQED2}
\begin{equation}\label{GlinearizedGamma0}
     G(\mathbf{k},\boldsymbol{\epsilon}_\mathbf{k})\big\vert_{k=\omega_0/c} = \sqrt{\frac{c^3\Gamma_\mu^{(0)}}{2\pi \omega_0^2}} A_\mu(\Omega_\mathbf{k},\pol{\mathbf{k}}).
\end{equation}
Here, we define $\Gamma_\mu^{(0)}$ as the recoil heating rate in the absence of squeezing, namely~\cite{GonzalezBallesteroPRA2019,MaurerQED2}
\begin{equation}\label{recoilMOTION}
    \Gamma_\mu^{(0)} = \frac{2\pi}{ c} \vert\alpha_0\vert^2 \left[\frac{\alpha}{2\epsilon_0(2\pi)^3}\right]^2 \omega_0^2 r_{0\mu}^2 \frac{8\pi k_0^4}{3}
   l_\mu,
\end{equation}
with  $l_\mu\equiv (1,2,7)\cdot \mathbf{e}_\mu/5$. The square-normalized function $A_\mu(\Omega_\mathbf{k},\pol{\mathbf{k}})$ describes the coupling strength between the motion along axis $\mu$ and the electromagnetic mode with polarization $\pol{\mathbf{k}}$ and propagation direction given by the two spherical angles $\Omega_{\mathbf{k}}$. It is given by
\begin{equation}\label{Ascattering}
    A_{\mu}(\Omega_\mathbf{k},\pol{\mathbf{k}})\equiv i e^{i\arg(\alpha_0)}\sqrt{\frac{3}{8\pi l_\mu}} (\boldsymbol{\varepsilon}_\mathbf{k}\cdot\mathbf{e}_x)\left[(\mathbf{e}_\mathbf{k}-\mathbf{e}_z)\cdot\mathbf{e}_\mu\right].
\end{equation}
The above function is 
directly related to the inelastic differential scattering cross section $d\sigma_\mu/d\Omega_\mathbf{k}$ (see Sec.~\ref{SectionIRP}) (see Sec. IIIB), and thus determines the angular distribution of the photons scattered by the motional degree of freedom $\mu$. For this reason, \eqnref{Ascattering} is deeply related to the amount of information about the $\mu$ degree of freedom that is contained in the light propagating along the direction $\Omega_\mathbf{k}$~\cite{TebbenjohannsPRA2019,MaurerQED2,supplementary}. This fact makes \eqnref{Ascattering} core to ground-state feedback cooling experiments~\cite{TebbenjohannsNature2021,MagriniNature2021}.

\subsection{Squeezing and collective mode basis}\label{SecHsqueezing}

Let us now consider the addition of squeezed light. We assume the squeezed light is produced by a standard cavity-enhanced spontaneous parametric down-conversion process (more details below), where the electromagnetic quantum correlations are modified in frequency domain~\cite{CollettPRA1984,BlowPRA1990}. It is thus convenient to perform a change of basis from plane-wave modes to a set of frequency-domain modes defined by the following annihilation operators,
\begin{align}
    \hat{a}_s(\omega) &\equiv \frac{k}{\sqrt{c}}\int d\Omega_\mathbf{k}\sum_{\pol{\mathbf{k}}} A_{s}(\Omega_\mathbf{k},\pol{\mathbf{k}}) \hat{a}(\keps),\\
    \hat{a}_j(\omega) &\equiv \frac{k}{\sqrt{c}}\int d\Omega_\mathbf{k}\sum_{\pol{\mathbf{k}}} A_{j}(\Omega_\mathbf{k},\pol{\mathbf{k}}) \hat{a}(\keps),
\end{align}
where $s$ and $j$ are a single and a continuous index, respectively. Each of the above modes is by construction a linear combination of plane waves with the same frequency $\omega$ and with a distribution of polarizations and propagation directions given by the weight functions $A_{s}(\Omega_\mathbf{k},\pol{\mathbf{k}})$ and $A_{j}(\Omega_\mathbf{k},\pol{\mathbf{k}})$, respectively. We single out the mode $\hat{a}_s(\omega)$, which we identify with the frequency-domain mode whose fluctuations will be modified by the squeezing. Thus, the function $A_{s}(\Omega_\mathbf{k},\pol{\mathbf{k}})$ describes the angular and polarization distribution of the squeezed light. Note that this function depends on the optical elements through which the squeezed light is sent before interacting with the particle, and thus can be tuned in each particular implementation. The squeezed mode $\hat{a}_s(\omega)$ is complemented by the set of modes $\hat{a}_j(\omega)$, chosen such that the set $\{\hat{a}_j(\omega),\hat{a}_s(\omega)\}$ forms a complete and orthonormal basis in the subspace of electromagnetic modes with frequency $\omega$, that is $[\hat{a}_\beta(\omega),\hat{a}_{\beta'}^\dagger(\omega')]=\delta_{\beta\beta'}\delta(\omega-\omega')$ for $\beta \in \{s,j\}$. We emphasize that the definition of these modes amounts to a change of basis
with
\begin{equation}
    \hat{a}(\keps) = \frac{\sqrt{c}}{k}\sum_{\beta=s,j}A_{\beta}^*(\Omega_\mathbf{k},\pol{\mathbf{k}})\hat{a}_\beta(\omega(k)),
\end{equation}
while the state of the electromagnetic field remains yet undefined. Orthonormality of this basis is guaranteed by choosing the weight functions to fulfill
\begin{equation}
    \int d\Omega_\mathbf{k}\sum_{\pol{\mathbf{k}}} A_{\beta}(\Omega_\mathbf{k},\pol{\mathbf{k}}) A_{\beta'}^*(\Omega_\mathbf{k},\pol{\mathbf{k}}) = \delta_{\beta\beta'}.
\end{equation}
The above equations also imply the reciprocal identity
\begin{equation}\label{reverseIdentity}
    \sum_{\beta=s,j}A^*_\beta(\Omega_{\mathbf{k}},\pol{\mathbf{k}})A_\beta(\Omega_{\mathbf{k}'},\pol{\mathbf{k}'}) = \delta(\Omega_{\mathbf{k}}-\Omega_{\mathbf{k}'})\delta_{\pol{\mathbf{k}}\pol{\mathbf{k}'}},
\end{equation}
(with the solid angle Dirac delta $\delta(\Omega_{\mathbf{k}}-\Omega_{\mathbf{k}'})\equiv \delta(\theta_{\mathbf{k}}-\theta_{\mathbf{k}'})\delta(\phi_{\mathbf{k}}-\phi_{\mathbf{k}'})$) which will be useful in the following.

Squeezing by parametric down-conversion produces a two-mode squeezed state in frequency domain between two sideband modes equally detuned from a laser carrier~\cite{CollettPRA1984,BlowPRA1990}, whose frequency we choose equal to the laser frequency $\omega_0$. To describe such squeezing in the modes $\hat{a}_s(\omega)$, we apply to the Hamiltonian the unitary transformation~\cite{BlowPRA1990}
\begin{equation}\label{Squeezingtrafo}
    \hat{T}_s(\eta) 
    \equiv 
    \exp
    \int \! d\omega \left[ \frac{\eta(\omega)}{2}\hat{a}_s^\dagger(\omega)\hat{a}_s^\dagger(2\omega_0-\omega)-\text{H.c.}
    \right]\!.
\end{equation}
The modulus and phase of the parameter $\eta(\omega) = \eta(2\omega_0-\omega) = r(\omega)e^{i\phi(\omega)}$ quantify the degree of squeezing and the squeezing phase of electromagnetic modes with frequency $\omega$, respectively (see Fig.~\ref{Figure1}b). To simplify integral expressions below, we assume the squeezed light is generated by an optical cavity with linewidth much smaller than $\omega_0$, which is usually fulfilled in typical experiments~\cite{VahlbruchPRL2016, MehmetOEx2011}. This is equivalent to assuming the squeezing parameter to be peaked around $\omega_0$, i.e. $r(0),r(2\omega_0)\ll r(\omega_0)$.  The squeezing transformation transforms the operator $\hat{a}_s(\omega)$ as
\begin{multline}
    \hat{T}_s(\eta) \hat{a}_s(\omega) \hat{T}_s^\dagger(\eta) = \hat{O}(\omega) \equiv
    \\
    \equiv \cosh [r(\omega)]\hat{a}_s(\omega)
    -e^{i\phi(\omega)}\sinh [r(\omega)]\hat{a}_s^\dagger(2\omega_0-\omega),
\end{multline}
and leaves the modes $\hat{a}_j(\omega)$ unaffected, as they remain in their vacuum state. Combing the above identities we express the Hamiltonian in the presence of squeezed light as
\begin{multline}\label{Hsqueezing}
 \frac{\hat{H}}{\hbar} = \sum_\mu\Omega_\mu\hat{b}^\dagger_\mu\hat{b}_\mu + \int d\omega \Delta(\omega) \sum_{\beta=s,j} \hat{a}^\dagger_\beta(\omega)\hat{a}_\beta(\omega) \\ 
 + \sum_\mu\hat{q}_\mu
 \int d\omega\left[\tilde{G}_{\mu s}(\omega) \hat{O}^\dagger(\omega)+ \text{H.c.}\right]\\ 
  +\sum_\mu\hat{q}_\mu
 \int d\omega\sum_{j}
\left[ \tilde{G}_{\mu j}(\omega)\hat{a}^\dagger_j(\omega) + \text{H.c.}\right],
\end{multline}
with $\Delta(\omega)\equiv \omega - \omega_0$, and with modified coupling rates
\begin{equation}
    \tilde{G}_{\mu \beta}(\omega) \equiv \frac{\omega}{\sqrt{c^3}}\int d\Omega_\mathbf{k}\sum_{\pol{\mathbf{k}}}
    A_{\beta}(\Omega_\mathbf{k},\pol{\mathbf{k}}) 
    G_{\mu}(\keps)\big\vert_{k=\omega/c}.
\end{equation}

In the following derivations it is convenient to use the above form for the Hamiltonian. However, note that by using the orthogonality relation \eqnref{reverseIdentity} one can rewrite the Hamiltonian \eqnref{Hsqueezing} in the form $\hat{H} = \hat{H}^{(0)} + \hat{H}'$, with $\hat{H}^{(0)}$ the Hamiltonian in the absence of squeezing, \eqnref{H0}.  The presence of squeezing thus modifies the light-matter interaction by adding an optomechanical coupling term that reads
\begin{multline}\label{extraH}
    \hat{H}' = \sum_\mu\hat{q}_\mu \int d\omega\tilde{G}_{\mu s}(\omega) \Big[ (\cosh[r(\omega)]-1)\hat{a}_s(\omega)
    \\
    -e^{i\phi(\omega)}\sinh[r(\omega)]\hat{a}_s^\dagger(2\omega_0-\omega)\Big]+ \text{H.c.}
\end{multline}
and which vanishes in the absence of squeezing, $r(\omega)=0$. This additional term in the Hamiltonian -- and thus all the modifications to the mechanical motion introduced by the squeezing -- depend only on the squeezing parameters, $r(\omega)$ and $\phi(\omega)$, and on the modified coupling rate, $\tilde{G}_{\mu s}(\omega)$. Even more, under the approximations undertaken in this work (see below), all relevant quantities depend only on their values at the laser frequency $\omega =\omega_0$. At this frequency the coupling rate can be written in the simple form
\begin{equation}\label{Gxi}
   \tilde{G}_{\mu s}(\omega_0) = \sqrt{\frac{\Gamma_\mu^{(0)}}{2\pi}} \xi_\mu,
\end{equation}
where we have defined the angular and polarization overlap between the squeezed light distribution and the angular and polarization distribution of the bare coupling rate,
\begin{equation}\label{eq:overlapIntegral}
    \xi_\mu \equiv \int d\Omega_\mathbf{k}\sum_{\pol{\mathbf{k}}} A_{s}(\Omega_\mathbf{k},\pol{\mathbf{k}})A_{\mu}(\Omega_\mathbf{k},\pol{\mathbf{k}}).
\end{equation}
Equations \eqref{extraH} and \eqref{Gxi} show that the impact of the squeezed light on the optomechanical interaction critically depends on the overlap integral \eqnref{eq:overlapIntegral}. We can understand this fact by noticing that this overlap integral $\xi_\mu$ quantifies how well the squeezed mode is matched to the optical mode that probes and drives the dynamics of the particle (the so-called ``interacting mode'', see Sec.~\ref{secDETECTION}).
In the extreme case where the distributions $A_{s}(\Omega_\mathbf{k},\pol{\mathbf{k}})$ and $A_{\mu}(\Omega_\mathbf{k},\pol{\mathbf{k}})$ are orthogonal, i.e. if one squeezes only the electromagnetic modes which do not interact with the motion along $\mu$,
Eq.~\eqref{Gxi} vanishes and the motion remains unaffected by the squeezing.

\section{Recoil heating rates and information radiation patterns}\label{secRecoilInfo}

In this section we derive the recoil heating rates of the mechanical motion, namely the rate of phonon increase due to scattering of laser photons (Sec.~\ref{subsecRecoilheating}), and the information radiation patterns, namely the angular and polarization distribution of such scattered photons (Sec.~\ref{SectionIRP}). Both these quantities are core to current ground-state cooling experiments based on optical detection of the particle motion~\cite{TebbenjohannsNature2021,MagriniNature2021,KambaOptExp2022}.

\subsection{Effective dynamics and recoil heating}\label{subsecRecoilheating}

To derive the effective dynamics of the center-of-mass motion under the influence of the squeezed free-space electromagnetic modes, we derive their dynamical equation by tracing out the electromagnetic field modes under the Born-Markov approximation~\cite{breuer2002theory,GonzalezBallesteroPRA2019}. This approximation is valid under two conditions. First, the optomechanical coupling in the presence of squeezing must be weak, i.e.,  $\vert G_{\mu}(\keps)\vert_{\omega=\omega_0} e^{r (\omega_0)}$ must be sufficiently small. In the absence of squeezing the coupling rates $\vert G_{\mu}(\keps)\vert_{\omega=\omega_0}$ fulfil this condition well, as the recoil heating rates calculated under this assumption agree with experimentally measured values~\cite{GonzalezBallesteroPRA2019,JainPRL2016}. Since, for the parameters considered in this work, the factor added by the squeezing remains of order one ($e^{ r (\omega_0)} \lesssim 6$), the weak coupling condition is safely fulfilled also in the presence of squeezing\footnote{
Note also the weak-coupling approximation is automatically fulfilled if the squeezed beam is tuned to reduce the recoil heating, regardless of the amount of squeezing, since at this configuration the total coupling rate to the squeezed mode is always reduced.}. Second, the bare coupling rates $\vert G_{\mu}(\keps)\vert$ must be smooth functions of $\omega$ around the laser frequency $\omega_0$, a condition that is also known to be fulfilled in free-space experiments.
Additionally, we assume the electromagnetic modes to be in a zero-temperature thermal state (vacuum), since the main contributions to the dynamics will stem from optical frequency modes. Note that decoherence induced by blackbody radiation is much weaker than photon recoil heating, and can thus be neglected~\cite{RomeroIsartPRA2011b,ChangPNAS2010}.

\begin{figure}[t!]
	\centering
	\includegraphics[width=\linewidth]{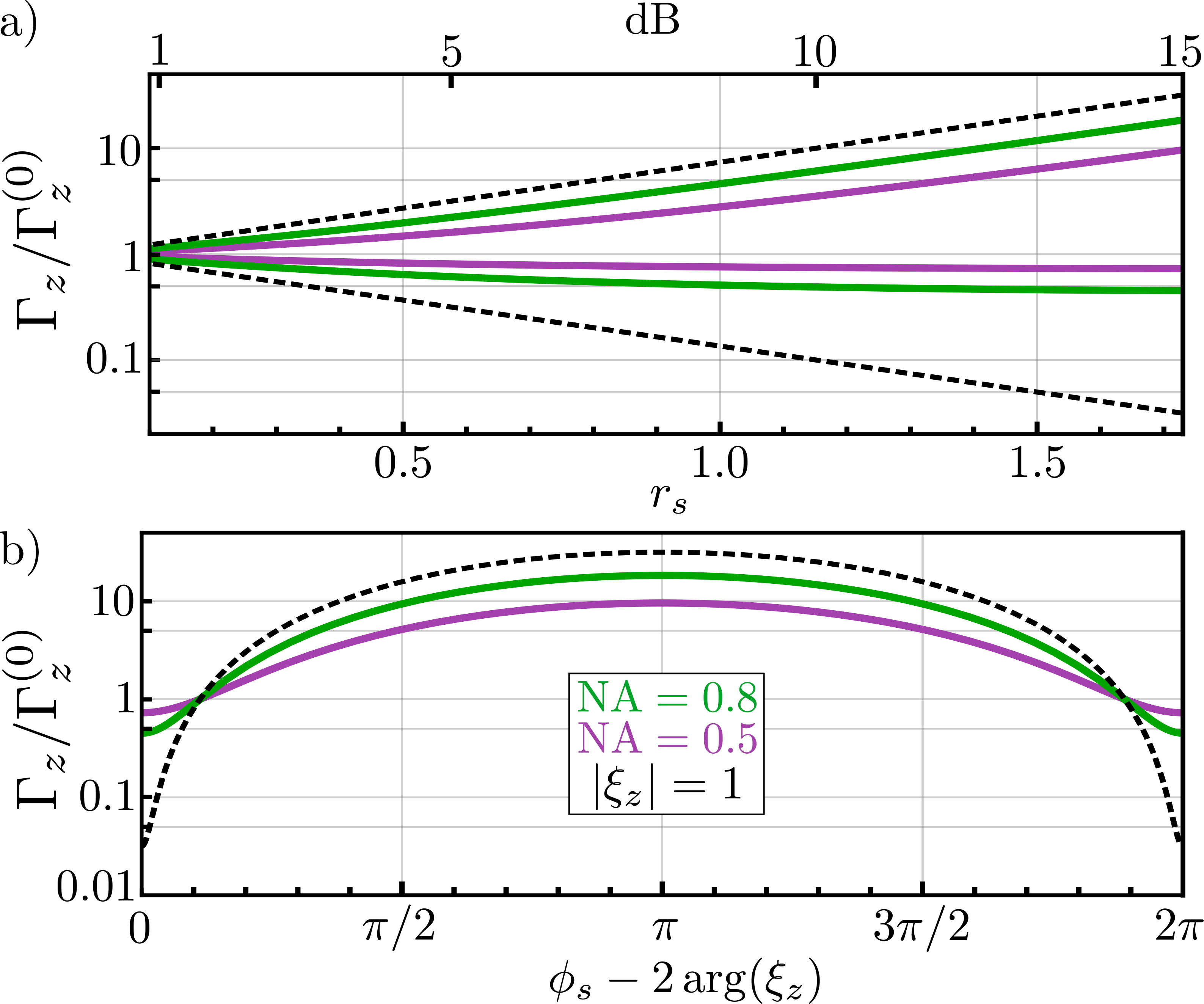}
	\vspace{-0.5cm}
	\caption{ a) Recoil heating rate for the mechanical motion along the $z$ axis as a function of the degree of squeezing for $\phi_s-2\arg(\xi_z)=0 $ (lower three lines) and $\phi_s-2\arg(\xi_z)=\pi$ (upper three lines). $\Gamma_z^{(0)}$ is the recoil heating rate in the absence of squeezing. b) Recoil heating rate versus squeezing phase for $r_s=1.73$ ($15$ dB). In both a) and b) we show the perfect overlap case ($\vert\xi_z\vert=1$, dashed) and the case of squeezed light propagating along the negative $z$ axis and focused by a lens of numerical apertures $0.8$ (green) and $0.5$ (purple), see text for details.
	}\label{Figure2}
	\vspace{-0.4cm}
\end{figure}

Under these approximations, we can trace out the electromagnetic degrees of freedom~\cite{breuer2002theory,GonzalezBallesteroPRA2019} to obtain a master equation for the mechanical density matrix, $\hat{\rho}$:
\begin{equation}\label{MasterEq}
    \frac{\text{d}\hat{\rho}}{\text{d}t} = \frac{1}{\im \hbar}\sum_\mu\left[\Omega_\mu\hat{b}^\dagger_\mu\hat{b}_\mu,\hat{\rho}\right] -\sum_{\mu\mu'}\frac{\Gamma_{\mu\mu'}}{2}\left[\hat{q}_\mu,\left[\hat{q}_{\mu'},\hat{\rho}\right]\right].
\end{equation}
Here, we have neglected the frequency shifts in the mechanical Hamiltonian. These shifts are independent of the squeezing and thus is the same as in free-space levitated optomechanics in the absence of squeezing, where it is known to be negligible. The dissipative rates are given by
\begin{multline}\label{crossrate}
    \Gamma_{\mu\mu'}=\Gamma_\mu^{(0)}\delta_{\mu\mu'} + 2\sqrt{\Gamma_\mu^{(0)}\Gamma_{\mu'}^{(0)}}\vert\xi_\mu\xi_{\mu'}\vert
    \\
    \times
    \left[s_0^2-s_0c_0\cos(\phi_s-\psi_\mu-\psi_{\mu'})\right].
\end{multline}
Here, we have defined the complex phase of the overlap \eqnref{eq:overlapIntegral} as $\psi_\mu\equiv\text{arg}(\xi_\mu)$, assumed $A_{s}(\Omega_\mathbf{k},\pol{\mathbf{k}}) \in \mathbb{R}$ for simplicity, and defined the squeezing functions $
    s_0\equiv \sinh(r_s)$ and $c_0\equiv \cosh(r_s)$, with $r_s = r(\omega_0 + \Omega_{\mu'})$ and $\phi_s = \phi(\omega_0 + \Omega_{\mu'})$. 
    For simplicity, noting that typical cavity-enhanced spontaneous down-conversion squeezed light sources have linewidths on the order of tens of MHz~\cite{VahlbruchPRL2016, MehmetOEx2011}, much larger than the mechanical frequencies $\Omega_\mu$, hereafter we approximate $r_s\approx r(\omega_0)$ and $\phi_s \approx \phi(\omega_0)$. 
    The off-diagonal terms $\mu\ne\mu'$ in \eqnref{crossrate} describe squeezing-mediated interactions between mechanical modes $\mu$ and $\mu'$. 
    Throughout this work we assume the overlap function is only significant along one direction ($\vert \xi_{\mu'\ne\mu}\vert \ll \vert \xi_\mu\vert \lesssim 1$), and hence these couplings can be neglected.
    Conversely, the diagonal terms $\Gamma_{\mu\mu}\equiv \Gamma_\mu$ correspond to the recoil heating rates for motion along the $\mu$-axis, which can be written as
    \begin{multline}\label{GammaRecoil}
        \frac{\Gamma_\mu}{\Gamma_\mu^{(0)}} = 1-\vert\xi_\mu\vert^2 
        \\
        \times\Big[1-e^{2r_s}\sin^2\frac{\phi_s-2\psi_\mu}{2}-e^{-2r_s}\cos^2\frac{\phi_s-2\psi_\mu}{2}\Big].
    \end{multline}
    The above expression shows that the squeezed vacuum state can modify the motional heating rates, an effect also predicted for atoms in the context of sideband cooling~\cite{AtomSqueezing1,AtomSqueezing2}.
    In the limit of no squeezing, $r_s\to0$, the heating rates Eq.~\eqref{GammaRecoil} recover the known expression~\eqnref{recoilMOTION}. According to \eqnref{GammaRecoil}, maximal reduction or increase of $\Gamma_\mu$ requires a squeezing phase $\phi_s=2\psi_\mu$ or $\phi_s=2\psi_\mu+\pi$ respectively, a large degree of squeezing, $r_s\gtrsim 1$, and perfect overlap between the squeezed mode distribution and the bare coupling distribution Eq.~\eqref{Ascattering} (related to the photon scattering amplitude in the absence of squeezing, see Sec.~\ref{SectionIRP}), i.e. $\xi_\mu \to 1$.
    
 In~\figref{Figure2}(a-b) we show the recoil heating rate as a function of the degree of squeezing $r_s$ and the phase of squeezing $\phi_s$, respectively. The black dashed lines correspond to perfect overlap $\vert\xi_\mu\vert=1$, and are independent of the motional axis considered. That is, 
the recoil heating along any of the three motional axes $\mu$ can be arbitrarily suppressed, by engineering the distribution of the squeezed mode $A_s(\Omega_\mathbf{k},\polK)$ such that $\xi_\mu \to 1$. This can be attained e.g. via wavefront shaping of the incoming squeezed light~\cite{HuepflArxiv2021,HuepflArxiv2022}.
For this perfect overlap scenario with a squeezing of 15dB ($r_s \approx 1.73$) -- achievable in spontaneous parametric down-conversion experiments at infrared laser wavelengths~\cite{VahlbruchPRL2016} -- the recoil heating rate can be suppressed by $\sim 98\%$. Note that squeezing the opposite quadrature results in an increase of recoil heating by a factor $\approx 32$.

Even without perfect mode overlap, the recoil heating rates can be significantly suppressed in current experiments. To show this, in~\figref{Figure2}(a-b) (colored lines) we show the recoil heating rate for the motion along the $\mu=z$ axis. For these plots we assume that the squeezed light is $x-$polarized and propagates along the negative $z$ axis (see \figref{Figure3}a), in order to maximize the overlap with the back-scattering dominated function $\vert A_z(\Omega_\mathbf{k},\polK)\vert^2$ (\figref{Figure3}b). In addition, we assume the squeezed light is focused onto the particle by a lens of numerical aperture NA, so that the squeezed mode distribution is Gaussian:
\begin{multline}\label{Gaussian}
    A_{s}(\Omega_\mathbf{k},\pol{\mathbf{k}}) = C e^{-(\sin\theta_\mathbf{k}/\text{NA})^2}\Theta(-\theta_\mathbf{k}+\pi/2)
    \\
    \times\left[-\delta_{\pol{\mathbf{k}},\mathbf{e}_{\theta_\mathbf{k}}}\cos\theta_\mathbf{k}\cos\phi_\mathbf{k}+\delta_{\pol{\mathbf{k}},\mathbf{e}_{\phi_\mathbf{k}}}\sin\phi_\mathbf{k}\right],
\end{multline}
with $\mathbf{e}_{\theta_\mathbf{k}}$ and $\mathbf{e}_{\phi_\mathbf{k}}$ the polar and azimuthal spherical vectors associated with the direction $\mathbf{k}\equiv k(\sin\theta_\mathbf{k}\cos\phi_\mathbf{k},\sin\theta_\mathbf{k}\sin\phi_\mathbf{k},\cos\theta_\mathbf{k})$, $\Theta(x)$ the Heaviside theta function, and $C$ a normalization constant. For such a Gaussian profile, and at $15$ dB squeezing, we have $\exp(-2r_s)\ll1$ and the reduction of the recoil heating rate is limited by the imperfect mode overlap, $\Gamma_z \approx \Gamma_z^{(0)}(1-\vert\xi_z\vert^2)$. As shown in~\figref{Figure2}(a-b), for NA$=0.9$ the recoil heating can be reduced by $\sim60\%$, or, at the opposite squeezing phase $\phi_s-2\psi_\mu=\pi$, increased  by a factor $\sim20$. These modifications should allow current motional reheating experiments~\cite{JainPRL2016} to clearly observe the impact of light squeezing in the measurement back-action noise.

\subsection{Information radiation patterns}\label{SectionIRP}

The information radiation patterns (IRP) describe the angular distribution of the photons that are scattered from the laser into other electromagnetic modes by absorption or excitation of a mechanical phonon. Since, as discussed below, these photons carry information about the particle position, knowledge of this distribution is key to optimal photodetector placement and thus to optimal feedback cooling. 

To calculate the IRP, it is sufficient to compute the total transition probability for the process where a phonon is excited in the mechanical mode $\mu$, initially in its ground state, via exchange of a photon between the laser mode and a free-space plane wave mode~\cite{MaurerQED2}. Since a squeezed vacuum has non-zero occupation we need to account for two possible processes, namely (i) the creation of a phonon by scattering a laser photon into the free-space mode $\{\keps\}$, with transition amplitude $\mathcal{T}_{\mu}^+(\keps)$,
and (ii) the creation of a phonon by absorbing a photon from the squeezed vacuum mode $\{\keps\}$ into the laser mode, with transition amplitude $\mathcal{T}_{\mu}^-(\keps)$. The transition amplitudes are given by
\begin{multline}
    \left[
    \begin{array}{c}
         \mathcal{T}_{\mu}^+(\keps)  \\
         \mathcal{T}_{\mu}^-(\keps) 
    \end{array}
    \right] \equiv \lim_{t_1, -t_0\to\infty}
    \\
    \bra{S}
    \left[
    \begin{array}{c}
         \alpha_0^*\hat{a}^\dagger(\keps)  \\
         \alpha_0\hat{a}(\keps) 
    \end{array}
    \right]
    \hat{b}_\mu\hat{U}(t_1,t_0)\ket{S},
\end{multline} 
with $\hat{U}(t_1,t_0)$ the time evolution operator for the linearized Hamiltonian  \eqnref{H0} and $\vert S \rangle \equiv \hat{T}_s(\eta)\vert 0\rangle_{\rm tot}$ the squeezed light state, with $\vert 0\rangle_{\rm tot}$ the compound vacuum state of the electromagnetic field and the mechanical motion.

Under the approximations discussed in Sec.~\ref{subsecRecoilheating}, we can calculate the above amplitudes to first order in time-dependent perturbation theory~\cite{MaurerQED2}, obtaining
\begin{equation}
    \mathcal{T}_{\mu}^\pm(\keps) =  \frac{if^\pm_{\mu}(\keps)}{2\pi k_0}\delta(k-k_0),
\end{equation}
in terms of the scattering amplitudes
\begin{multline}\label{fplus}
     f_{\mu}^+(\keps)=-    \alpha_0^*\sqrt{\frac{(2\pi)^3\Gamma_\mu^{(0)}}{c}}
     \\
     \times\left[ A_{\mu}(\Omega_\mathbf{k},\pol{\mathbf{k}})+
     A_{s}^*(\Omega_\mathbf{k},\pol{\mathbf{k}})
   g_\mu\right],
\end{multline}
\begin{equation}
\label{fminus}
     f_{\mu}^-(\keps)=- \alpha_0\sqrt{\frac{(2\pi)^3\Gamma_\mu^{(0)}}{c}}
     A_{s}^*(\Omega_\mathbf{k},\pol{\mathbf{k}})
     g_\mu^*,
\end{equation}
and the squeezing coefficient $g_\mu \equiv  \xi_\mu  s_0(s_0 - c_0 \exp[i(\phi_s-2\psi_\mu)])$. These scattering amplitudes can be used to define the differential scattering cross section~\cite{Cohen2004/2,MaurerQED2}
\begin{equation}
    \frac{d\sigma_\mu}{d\Omega_\mathbf{k}}=\sum_{\pol{\mathbf{k}}}\left(\vert f_{\mu}^+(\keps)\vert^2-\vert f_{\mu}^-(\keps)\vert^2\right).
\end{equation}
Note that to describe the total probability of scattering photons into mode $\{\keps\}$ the two scattering amplitudes must be subtracted, as the second process removes photons from the mode $\{\keps\}$ and transfers them into the laser mode. The information radiation pattern is defined as the normalized differential scattering cross section,
\begin{equation}
    \mathcal{I}_\mu(\Omega_\mathbf{k})\equiv \frac{(d\sigma_\mu/d\Omega_\mathbf{k})}{\int d\Omega_{\mathbf{k}}(d\sigma_\mu/d\Omega_\mathbf{k})} = \frac{c}{(2\pi)^3\vert\alpha_0\vert^2\Gamma_\mu} \frac{d\sigma_\mu}{d\Omega_\mathbf{k}}.
\end{equation}

\begin{figure}[t!]
	\centering
	\includegraphics[width=\linewidth]{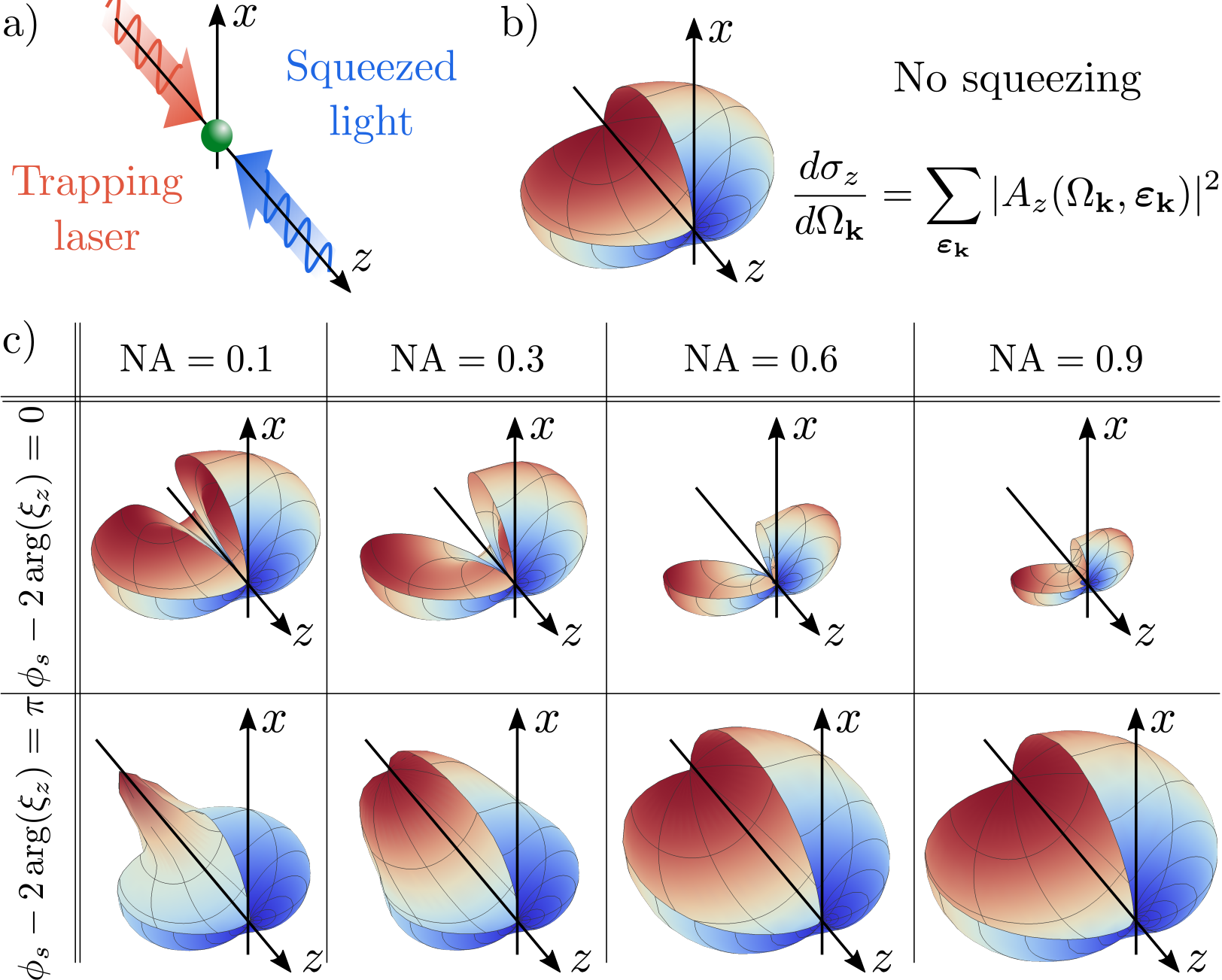}
	\caption{a) Modification of the recoil heating of the motion along the $z$-axis is maximized for a squeezed beam counter-propagating with respect to the  light.
	b) Differential scattering cross section $d\sigma_z/d\Omega_\mathbf{k}$ in the absence of squeezing (information radiation pattern). The value of the function is encoded in the radial distance to the origin and the colour scale encodes the relative amplitude within each panel.
	c) Differential scattering cross section for squeezed light with a Gaussian spatial profile of different numerical apertures. Upper and lower rows correspond to $13$dB and $\phi_s- 2\arg(\xi_z)=0$ and to $3$dB and $\phi_s- 2\arg(\xi_z)=\pi$, respectively.}\label{Figure3}
	\vspace{-0.2cm}
\end{figure}

In \figref{Figure3}(b) we show the differential scattering cross section for the $\mu=z$ mechanical mode in the absence of squeezing. As shown by the figure and known in the literature~\cite{TebbenjohannsPRA2019,TebbenjohannsNature2021,MagriniNature2021,MaurerQED2,AndreiArxiv2022}, the photon scattering in this case is dominated by back-scattering. In \figref{Figure3}(c) we show the differential scattering cross section in the presence of a counter-propagating squeezed beam with Gaussian intensity profile, \eqnref{Gaussian}. At $\phi_s = 2\psi_z$, the total amount of inelastically scattered light is suppressed, whereas at $\phi_s = 2\psi_z+\pi$ it is enhanced. This is consistent with the results of~\figref{Figure2}(a-b), that is: a decrease ($\phi_s = 2\psi_z$) or an increase ($\phi_s = 2\psi_z + \pi$) in the amount of information about the motion that is leaked to propagating electromagnetic modes results in a decrease or increase of the total mechanical decoherence. As the squeezed beam becomes more focused (increase in NA) the overlap $\xi_z$ increases, efficiently suppressing photon scattering across a wider solid angle. In the limit of perfect overlap ($\xi_\mu\to 1$) and high squeezing ($r_s\gg 1$) we obtain $d\sigma_\mu/d\Omega_\mathbf{k} \to 0$ and the mechanical motion is perfectly protected from laser shot noise decoherence.

\section{Position detection beyond the standard quantum limit}\label{secDETECTION}

In this section we model the optical detection of the mechanical motion in the presence of squeezed light. We first compute the interacting modes and their input-output relations in~\secref{inputoutputSEC}. Then, we compute the minimum detectable signal as a function of squeezing parameters in~\secref{SminSEC}.

\subsection{Interacting modes and input-output theory}\label{inputoutputSEC} 

Our aim is to derive the state of the electromagnetic field after its interaction with the mechanical motion.  We start by deriving, from the Hamiltonian~\eqnref{Hsqueezing}, the coupled Heisenberg dynamical equations for the mechanical and photon operators,
\begin{multline}\label{EOMb}
 \frac{d\hat{b}_\mu}{dt} = -i\Omega_\mu\hat{b}_\mu -i \int d\omega\\  \bigg\lbrace \sum_j \tilde{G}_{\mu j}(\omega)\hat{a}^\dagger_j(\omega)
 +\tilde{G}_{\mu s}(\omega) \hat{O}^\dagger(\omega) 
 +  \text{H.c.}\bigg\rbrace,
\end{multline}
\begin{equation}\label{EOMabeta}
    \frac{d\hat{a}_\beta(\omega)}{dt}=-i\Delta(\omega)\hat{a}_\beta(\omega)-i\sum_\mu\hat{q}_\mu Q_{\mu \beta}(\omega),
\end{equation}
with $Q_{\mu\beta} (\omega)= \tilde{G}_{\mu j}(\omega)$ for $\beta=j$ and $Q_{\mu\beta}(\omega) = \tilde{G}_{\mu s}(\omega)\cosh[r(\omega)] - \tilde{G}^*_{\mu s}(2\omega_0-\omega)\sinh[r(\omega)] e^{i\phi(\omega)}$ for the squeezed mode $\beta=s$. We now formally solve Eqs.~\eqref{EOMabeta} in terms of an initial condition, namely a time  $t_0 \to -\infty$ in the distant past,
\begin{multline}\label{ajformal}
    \hat{a}_\beta(\omega,t)=\hat{a}_\beta(\omega,t_0)e^{-i\Delta(\omega)(t-t_0)}
    \\-i\int_{t_0}^tds\sum_\mu\hat{q}_\mu(s)e^{-i\Delta(\omega)(t-s)}Q_{\mu \beta}(\omega),
\end{multline}
and introduce the result into \eqnref{EOMb}. Under the approximations undertaken in this work we can cast the resulting equation in the form\footnote{
Here we also assume that $\{r(\omega),\phi(\omega)\}$ evolve slowly around $\omega_0$, which is true in typical experiments as discussed above.
}
\begin{equation}\label{EOMboperator}    
    \frac{d\hat{b}_\mu}{dt}=-i\Omega_\mu\hat{b}_\mu-i\sqrt{\Gamma^{(0)}}\left[\tilde{a}_\mu^{\rm in} (t)
    +\text{H.c.}\right].
\end{equation}
The ladder operators $\tilde{a}_\mu^{\rm in} (t)$ fulfil bosonic commutation relations, $[\tilde{a}_\mu^{\rm in} (t),\tilde{a}_{\mu'}^{\text{in}\dagger} (t')]=\delta_{\mu\mu'}\delta(t-t')$. They correspond to the ``interacting mode''~\cite{MagriniNature2021,AndreiArxiv2022}, namely the single collective electromagnetic mode that couples to the mechanical degree of freedom $\mu$. This operator reads
\begin{multline}\label{interactingmode}
        \tilde{a}_\mu^{\rm in} (t) \equiv \tilde{a}_\mu^{\text{in} (0)} (t) + \left(1/\sqrt{\Gamma^{(0)}}\right)\int d\omega e^{-i\Delta(\omega)(t-t_0)} \\ \times   
        \tilde{G}_{\mu s}^*(\omega)\left[\hat{O}(\omega,t_0)-\hat{a}_s(\omega,t_0)\right]
       ,
\end{multline}
with the interacting mode operator in the absence of squeezing defined as~\cite{MagriniNature2021,AndreiArxiv2022}
\begin{multline}
        \tilde{a}_\mu^{\text{in} (0)} (t)\equiv \frac{1}{\sqrt{\Gamma^{(0)}}}\int d\mathbf{k}\sum_{\boldsymbol{\varepsilon}_\mathbf{k}} G_\mu^*(\mathbf{k},\boldsymbol{\varepsilon}_\mathbf{k})\allowbreak \\
        \times\hat{a}(\mathbf{k},\boldsymbol{\epsilon}_\mathbf{k},t_0)e^{-i\Delta(\omega_k)(t-t_0)}.
\end{multline}

Since the interacting mode is the only electromagnetic mode interacting with the mechanical degree of freedom, it contains all the information about the mechanical motion that carried by the light. The next step is thus to derive input-output relations for the interacting mode. For this purpose, we re-derive Eqs.~\eqref{ajformal}-\eqref{EOMboperator} in terms of a final condition, namely a time $t_1\to \infty$ in the distant future. This allows us to define output operators $\tilde{a}^{\rm out}_\mu (t)$, given by \eqnref{interactingmode} under the substitution $t_0\to t_1$, and to formally relate them to the input operators,
\begin{equation}\label{inputoutputX}
    \tilde{X}_\mu^{\rm out}(t)=\tilde{X}_\mu^{\rm in}(t),
\end{equation}
\begin{multline}\label{inputoutputY}
    \tilde{Y}_\mu^{\rm out}(t) = \tilde{Y}_\mu^{\rm in}(t)-\sqrt{4\Gamma_\mu^{(0)}}\hat{q}_{\mu}(t)
    \\
    = \tilde{Y}_\mu^{\rm in}(t) -\sqrt{4\Gamma_\mu^{(0)}}\hat{q}_{\mu}^f(t)-\sqrt{4\Gamma_\mu^{(0)}}\hat{q}_{\mu}^{\rm ba}(t),
\end{multline}
where we have defined the interacting mode quadratures $\tilde{X}_\mu^{\rm in /out}(t) \equiv \tilde{a}_\mu^{\text{in/out}, \dagger} (t) + \text{H.c.}$ and $\tilde{Y}_\mu^{\rm in /out}(t) \equiv i\tilde{a}_\mu^{\text{in/out}, \dagger} (t) + \text{H.c.}$ In the last step of~\eqnref{inputoutputY} we have used the formal solution of~\eqnref{EOMboperator}, namely
\begin{multline}\label{bmuformal}
    \hat{b}_\mu(t)=
    \hat{b}_\mu(t_0)e^{-i \Omega_\mu(t-t_0)} \\ -i \sqrt{\Gamma_\mu^{(0)}}\int_{t_0}^tds \tilde{X}_\mu^{\rm in}(s)e^{-i \Omega_\mu (t-s)}
    \equiv \hat{b}_\mu^{f}(t)+\hat{b}_\mu^{\rm ba}(t),
\end{multline}
to define two contributions, namely the unperturbed, free evolution of the mechanical mode $\hat{b}_\mu^f(t)\equiv \hat{b}_\mu(t_0)\exp[-i \Omega_\mu(t-t_0)]$, which is the signal we aim at detecting, and its modification due to the interaction with the light, which constitutes the back-action term $\hat{b}_\mu^{\rm ba}(t)$.

\subsection{Minimum detectable signal as a function of squeezing}\label{SminSEC}

To model photodetection we compute the output power spectral densities, defined for two arbitrary quadratures $A_\mu,B_\mu\in \{\tilde{X}^{\rm in/out}_\mu,\tilde{Y}^{\rm in/out}_\mu\}$
as
\begin{equation}
    S_{\mu, AB}(\omega)\equiv  \int \frac{d\tau}{2\pi} e^{i\omega\tau}\langle\hat{A}_\mu(t+\tau)\hat{B}_\mu(t)\rangle,
\end{equation}
with the expectation value taken over the vacuum state. As shown by Eqs.~\eqref{inputoutputX}-\eqref{inputoutputY}, optimal photodetection of the mechanical motion relies on detecting the quadrature $\tilde{Y}^{\rm out}(t)$. Its power spectral density, assuming no initial correlations between the electromagnetic field and the mechanical motion ($\langle \hat{q}_\mu^f(t_0+\tau)\hat{a}_\kappa(t_0)\rangle = 0$), reads
 \begin{multline}\label{SPPout}
    S_{\mu, YY}^{\rm out} (\omega) = S_{\mu, YY}^{\rm in}  +\frac{4\Gamma_\mu^{(0)}}{r_{0\mu}^2}S_{\mu}^f(\omega)
    \\+\frac{4\Gamma_\mu^{(0)}}{r_{0\mu}^2} S_{\mu}^{\rm ba}(\omega)
    - \frac{\sqrt{4\Gamma_\mu^{(0)}}}{r_{0\mu}} S_{\mu}^{\rm c}(\omega).
\end{multline}
The first term in \eqnref{SPPout} describes the original fluctuations of the light field, namely the original imprecision in the light due to photon shot noise~\cite{TebbenjohannsPRA2019}.
The second term contains the unperturbed PSD of the mechanical mode,
\begin{equation}
    S_{\mu}^\text{f}(\omega)\equiv\frac{r_{0\mu}^2}{2\pi}\int_{\mathbb{R}}\text{d}\tau \langle\hat{q}^f_\mu(t+\tau)\hat{q}^f_\mu(t)\rangle e^{i \omega\tau},
\end{equation}
and corresponds to the signal to be detected in the output light. The third term in Eq.~\eqref{SPPout} describes the disturbance in the measurement caused by the back-action of the light on the mechanical mode~\cite{TebbenjohannsPRA2019}. Finally, the fourth term describes the amplitude-phase correlations of the output light. Both terms can be explicitly calculated in terms of power spectral densities of the input light. Specifically, as shown in Appendix~\ref{AppPSD},
\begin{multline}\label{Sba}
     S_{\mu}^\text{ba}(\omega)\equiv\frac{r_{0\mu}^2}{2\pi}\int_{\mathbb{R}}\text{d}\tau \langle\hat{q}^\text{ba}_{\mu}(t+\tau)\hat{q}^\text{ba}_{\mu}(t)\rangle e^{i \omega\tau}
     \\ =r_{0\mu}^2\Gamma_\mu^{(0)}S_{\mu, XX}^{\rm in}\vert 2m\Omega_{\mu}\chi_{\mu}(\omega)\vert^2
     ,
\end{multline}
and
\begin{multline}\label{Sbacrossed}
   S^c_\mu(\omega) \equiv\frac{r_{0\mu}}{2\pi}\int_\mathds{R} d\tau e^{i\omega\tau}
   \\
   \times \bigg[\langle\tilde{Y}_\mu^{\rm in}(t+\tau)\hat{q}_{\mu}^{\rm ba}(t)\rangle +
   \langle\hat{q}_{\mu}^{\rm ba}(t+\tau)\tilde{Y}_\mu^{\rm in}(t)\rangle\bigg] 
   \\=r_{0\mu}\sqrt{\Gamma_\mu^{(0)}} \text{Re}[2m\Omega_{\mu}\chi_{\mu}(\omega)]
   \left(S_{\mu,XY}^{\rm in}+S_{\mu,YX}^{\rm in}\right),
\end{multline}
where we define the mechanical susceptibility
\begin{equation}
	\chi_{\mu}(\omega)\equiv\frac{1}{m}\frac{1}{\Omega_{\mu}^2-\omega^2-i\gamma_\mu\omega} \equiv \frac{1}{m\Omega_\mu^2}\tilde{\chi}_\mu(\omega),
\end{equation}
with $\gamma_\mu$ the mechanical damping, due to e.g. collisions of the particle with the surrounding gas, and $\tilde{\chi}_\mu(\omega)$ an adimensional susceptibility defined for convenience.
The power spectral densities of the input quadratures, which reflect the modified fluctuations present in the initial squeezed state, are given in Appendix~\ref{AppPSD}.

\begin{figure}[t!]
	\centering
	\includegraphics[width=\linewidth]{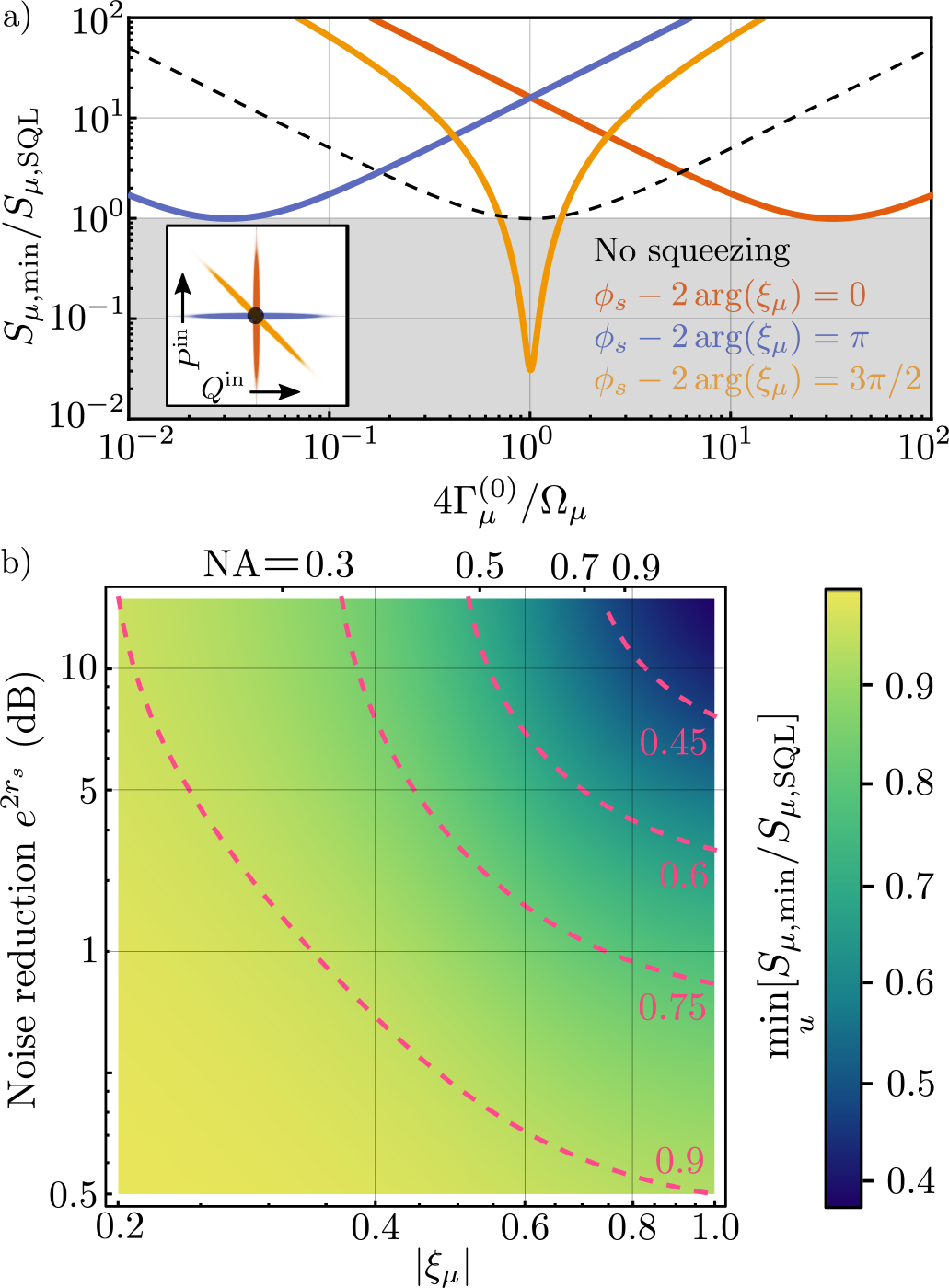}
	\caption{a) Minimum detectable signal for $\omega \ll \Omega_\mu$ in units of its SQL as a function of $4\Gamma^{(0)}_\mu/\Omega_\mu$, in the absence of squeezing (dashed) and for $15$ dB squeezing and perfect overlap, $\vert\xi_\mu\vert=1$, along different quadratures (colored). Inset: Wigner function of the input light for these curves (see Appendix~\ref{AppWigner} for details). The black circle indicates the Wigner function size in the absence of squeezing.
	b) Minimum signal $\min_u(S_{\mu,\rm min}/S_{\mu,\rm SQL})\vert_{\phi_s-2\arg(\xi_\mu)=3\pi/2}$ as a function of degree of noise reduction $e^{2r_s}$ and mode overlap $\vert \xi_\mu\vert$. The upper ticks mark the value of the overlap $\vert\xi_z\vert$ corresponding to a squeezed beam with Gaussian angular profile and NA$=0.3$, $0.5$, $0.7$, and $0.9$. Constant-value isolines are marked by dashed curves.
 } \label{figure4}
    \end{figure}

    \begin{figure*}[t!]
	\centering
	\includegraphics[width=\linewidth]{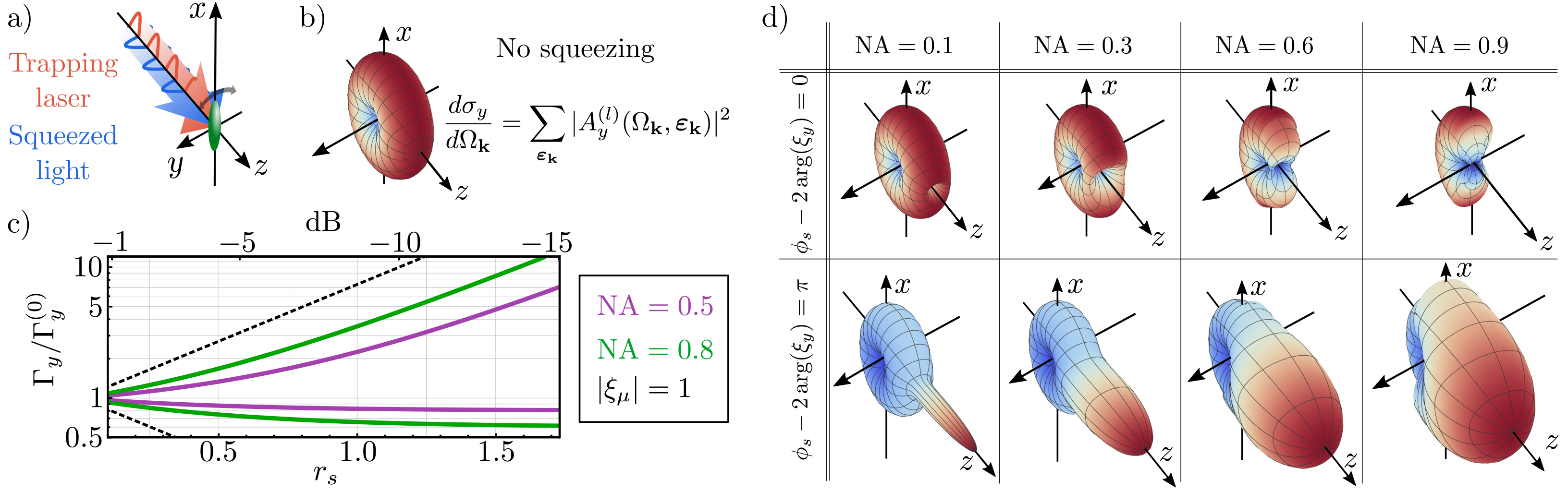}
  \vspace{-0.5cm}
	\caption{a) Our formalism can be directly extended to the libration of an anisotropic particle around its equilibrium orientation along the laser polarization. b) Differential scattering cross section for the libration along $y$ in the absence of squeezing. c) Recoil heating for the libration along $y$ as a function of degree of squeezing, for $\phi_s-2\arg(\xi_y) = 0$ (upper three lines) and $\phi_s-2\arg(\xi_y) = \pi$ (lower three lines). The black dashed curves correspond to the perfect overlap case $\vert\xi_y\vert=1$ and are identical to the black dashed lines in \figref{Figure2}, whereas colored curves to a squeezed beam focused by a lens of numerical apertures NA$=0.5$ (purple) and NA$=0.8$ (green). d) Differential scattering cross section for the libration along $y$, for a co-propagating squeezed light beam (see panel a) focused by lenses of different numerical apertures. Upper and lower rows correspond to $13$dB and $\phi_s- 2\arg(\xi_\mu)=0$ and to $3$dB and $\phi_s- 2\arg(\xi_\mu)=\pi$, respectively.
 \vspace{-0.3cm}
 } \label{figure5}
    \end{figure*}

To detect the mechanical motion in the output light, the second term in Eq.~\eqref{SPPout} (the signal) must be larger than the sum of the remaining terms (imprecision, back-action, and correlations). We define the minimum detectable mechanical signal as 
\begin{equation}
    S_{\mu,\rm min}(\omega) \equiv   \frac{r_{0\mu}^2}{4\Gamma_\mu^{(0)}} S_{\mu, YY}^{\rm in}  + S_{\mu}^{\rm ba}(\omega)-\frac{r_{0\mu}}{\sqrt{4\Gamma_\mu^{(0)}}}S^c_\mu(\omega).
\end{equation}
Using all the above expressions we can cast it as
\begin{multline}\label{Sminimumgeneral}
    S_{\mu,\rm min}(\omega)= \frac{\pi S_{\mu,\rm SQL}(\omega)}{2}\bigg[u \vert \tilde{ \chi}_{\mu}(\omega)\vert S_{\mu, XX}^{\rm in} \\+ \frac{S_{\mu, YY}^{\rm in}}{u \vert \tilde{ \chi}_{\mu}(\omega)\vert} -\frac{\text{Re}(\tilde{ \chi}_{\mu}(\omega)) }{ \vert \tilde{ \chi}_{\mu}(\omega)\vert}(S_{\mu, XY}^{\rm in} +S_{\mu, YX}^{\rm in})\bigg].
\end{multline}
Here, $u \equiv 4\Gamma_\mu^{(0)}/\Omega_\mu$ is the adimensional ratio between the bare recoil heating rate (i.e. in the absence of squeezing) and the mechanical frequency, and $S_{\mu,\rm SQL}\equiv  \vert \tilde{ \chi}_{\mu}(\omega)\vert r_{0\mu}^2/(\pi\Omega_\mu)$ is the absolute minimum signal that can be detected in the absence of squeezing, namely the Standard Quantum Limit. Indeed, in the absence of squeezing  ($2\pi S_{\mu, XX}^{\rm in}, 2\pi S_{\mu, YY}^{\rm in}\to 1$ and $S_{\mu, XY}^{\rm in}+S_{\mu, YX}^{\rm in}\to 0$) Eq.~\eqref{Sminimumgeneral} reduces to 
\begin{equation}
    S_{\mu,\rm min}^{(0)} = \frac{S_{\mu,\rm SQL}}{2} \left[u  \vert \tilde{ \chi}_{\mu}(\omega)\vert +\frac{1}{u  \vert \tilde{ \chi}_{\mu}(\omega)\vert}\right],
\end{equation}
which, as a function of $u$, has a minimum value $\min_u (S_{\mu,\rm min})=S_{\mu,\rm SQL}(\omega)$ at $u= \vert \tilde{ \chi}_{\mu}(\omega)\vert^{-1}$. In the presence of squeezing, \eqnref{Sminimumgeneral} reaches a different minimum value at $u= \vert \tilde{ \chi}_{\mu}(\omega)\vert^{-1}[S_{\mu, YY}^{\rm in}/S_{\mu, XX}^{\rm in}]^{1/2}$,
which is given by
\begin{multline}\label{Sabsolutemin}
    \min_u\left[\frac{S_{\mu,\rm min}(\omega)}{S_{\mu,\rm SQL}(\omega)}\right]
    = \sqrt{\left(2\pi S_{\mu, XX}^{\rm in}\right)\left(2\pi S_{\mu, YY}^{\rm in}\right)}
        \\
        -\frac{\text{Re}(\tilde{ \chi}_{\mu}(\omega)) }{ \vert \tilde{ \chi}_{\mu}(\omega)\vert}2\pi\frac{S_{\mu, XY}^{\rm in} +S_{\mu, YX}^{\rm in}}{2}.
\end{multline}
At resonance, $\omega=\Omega_{\mu}$, the above expression is always larger than $1$, and the minimum signal remains above the Stardard Quantum Limit even in the presence of squeezing. For $\omega \neq\Omega_{\mu}$, Eq.~\eqref{Sabsolutemin} is minimized at a squeezing angle
 $\phi_s-2\psi_{\mu}=3\pi/2$ if $\text{Re}(\tilde{ \chi}_{\mu}(\omega))>0$ (or equivalently, for $\omega<\Omega_\mu$), and at a squeezing angle $\phi_s-2\psi_{\mu}=\pi/2$ if $\text{Re}(\tilde{ \chi}_{\mu}(\omega))<0$ (or equivalently, for $\omega>\Omega_\mu$). In both cases the minimum detectable signal can be written in compact form as
\begin{multline}
    \min_{u,\phi_s}\left[\frac{S_{\mu,\rm min}(\omega)}{S_{\mu,\rm SQL}(\omega)}\right]=1-\vert\xi_\mu\vert^2
    \\
    +\vert\xi_\mu\vert^2\sum_{\eta=\pm 1}\frac{e^{2 \eta r_s}}{2}\left[1-\eta\frac{\vert\text{Re}(\tilde{ \chi}_{\mu}(\omega))\vert }{ \vert \tilde{ \chi}_{\mu}(\omega)\vert}\right]
\end{multline}
The sensitivity is thus maximized 
 for $\vert\text{Re}(\tilde{ \chi}_{\mu}(\omega))/ \tilde{ \chi}_{\mu}(\omega)\vert \to 1$ or, equivalently for off-resonant frequencies $\vert\omega-\Omega_{\mu}\vert \gg\gamma_\mu$. This condition is fulfilled at almost all frequencies in typical levitated optomechanics experiments where $\gamma_\mu\ll\Omega_\mu$. Hereafter we focus on this range of frequencies.

    In~\figref{figure4}(a) we show $S_{\mu,\rm min}$ in the presence of squeezing and in the low frequency limit $\omega \ll \Omega_\mu$ for simplicity. For $\phi_s=2\psi_\mu$ the back-action, and thus the recoil heating rate, is reduced (see Fig.~\ref{Figure2}), and the minimum signal shifts to a higher value of $u$. For $\phi_s=2\psi_\mu+\pi$ the opposite behavior occurs. At $\phi_s=2\psi_\mu+3\pi/2$ the minimum detectable signal decreases below the SQL. At this squeezing phase, the amplitude-phase correlations $S_\mu^c(\omega)$ in the input light minimize the detection noise~\cite{bowen2015quantum,AndreiArxiv2022,MagriniPRL2022}, and the sensitivity reads
    \begin{equation}\label{minuS}
       \min_u\left[\frac{S_{\mu,\rm min}(\omega)}{S_{\mu,\rm SQL}(\omega)}\right]_{\phi_s-2\psi_\mu=3\pi/2} = 1-\vert\xi_\mu\vert^2\left(1-e^{-2r_s}\right).
    \end{equation}
    This expression, shown in~\figref{figure4}b, is identical to the minimum achievable value of $\Gamma_\mu/\Gamma_\mu^{(0)}$ (see~\eqnref{GammaRecoil}). The results of~\figref{figure4} extend the concept of enhancement of displacement sensitivity via injection of squeezed vacuum light, studied extensively in 1D- and cavity optomechanics in contexts such as gravitational wave detection~\cite{CavesPRD1981,Unruh1983,BondurantPRD1984,Jaekel1990,PacePRA1993,bowen2015quantum,KimblePRD2001}, to levitated optomechanics in free-space, i.e. without optical cavities. We remark that, in contrast to methods where squeezed light is used only at the detector (e.g. injected at the vacuum port of a detection interferometer), the sensitivity enhancement shown in~\figref{figure4} stems from the fundamental modification of light-motion interaction attained by shining squeezed light directly at the nanoparticle. Note that this enhancement can also be used to overcome technical limitations in the optical measurement of the mechanical displacement, which unavoidably reduce the detection efficiency below unity. Our results show that, by properly tuning the squeezing parameters, this efficiency can be effectively increased, which can lead to performance improvements for measurement-based cooling schemes~\cite{SchafermeierNature2016}.

\section{Extension to libration of nanorotors}\label{secROTORA}

Let us show how the formalism derived in this work can be directly extended from the motional degrees of freedom of spherical particles to the librational degrees of freedom of nanorotors~\cite{HoangPRL2016,KuhnNatComm2017,AhnPRL2018,AhnNatNano2020,BangPRR2020,DelordNature2020,vanderLaanPRA2020,vanderLaanPRL2021}. We consider an anisotropic, cylindrically symmetric, sub-wavelength dielectric particle in the presence of a laser beam propagating along the positive $z$ axis and polarized along the $x$ axis (Fig.~\ref{figure5}(a)). The particle is assumed fixed at the origin but free to rotate along its axes, and has a polarizability tensor $\overline{\boldsymbol{\alpha}}$ which, in the body frame, reads $\overline{\alpha}_{ij} = \alpha_i\delta_{ij}$ with $\alpha_1=\alpha_2<\alpha_3$. The laser beam aligns the long axis of the particle along the polarization axis $x$, allowing it to undergo small angular displacements $\eta_\mu$ along the axes $\mu=\{y,z\}$.  As shown in Appendix~\ref{AppA}, the Hamiltonian of the electromagnetic and librational degrees of freedom has the same form as \eqnref{H0}, namely
\begin{multline}\label{Htransformed2MAIN}
    \hat{H}^{(0)}/\hbar= \sum_\mu\Omega_\mu\hat{b}^\dagger_\mu\hat{b}_\mu + \hat{H}_{\rm EM,r}/\hbar
    \\+ \int d\mathbf{k}\sum_{\boldsymbol{\varepsilon}_\mathbf{k}}\sum_{\mu}\hat{q}_\mu\left(G_{\mu}^{(l)}(\keps)\hat{a}^\dagger(\keps) +\text{H.c.}\right),
\end{multline}
with linearized coupling rates for libration $G_{\mu}^{(l)}$ (labeled by the super-index $(l)$ to distinguish from their center-of-mass motion analogues) which, at the laser frequency $\omega=\omega_0$, take the form
of~\eqnref{GlinearizedGamma0}, i.e.,
\begin{equation}
    G_{\mu}^{(l)}(\keps)\big\vert_{k=k_0} = \sqrt{\frac{c^3\Gamma_\mu^{(0)}}{2\pi \omega_0^2}} A_{\mu}^{(l)}(\Omega_\mathbf{k},\pol{\mathbf{k}}),
\end{equation}
where $\Gamma_\mu^{(0)}$ is the recoil heating rate in the absence of squeezing~\cite{SebersonPRA2020},
\begin{equation}\label{recoilROTATIONmain}
    \Gamma_\mu^{(0)} = \frac{V}{4\pi^2c}\vert \alpha_0\vert^2\left(\frac{\alpha_3-\alpha_1}{2\epsilon_0(2\pi)^3}\right)^2\frac{8\pi k_0^2}{3}r_{0\mu}^2\omega_0^2,
\end{equation}
 and where the orthonormal and square-normalized libration angular distributions are given by
\begin{equation}
    A_{\mu}^{(l)}(\Omega_\mathbf{k},\pol{\mathbf{k}})\equiv -e^{i\text{arg}(\alpha_0)}\sqrt{\frac{3}{8\pi}}(\pol{\mathbf{k}}\cdot\mathbf{e}_\mu).
\end{equation}
Note that these functions are simpler than their motional analogues. Specifically, in the absence of squeezing the differential scattering cross sections for libration along the $\mu$ axis take the form of the field radiated by a dipole pointing along the $\mu$ axis (Fig.~\ref{figure5}(b)).

Once expressed in this form, the whole derivation in all the sections above remains valid as it only assumes that the form of the Hamiltonian is given by~\eqnref{H0} and that the coupling rates fulfill the identity~\eqnref{GlinearizedGamma0}. All the expressions in the article are thus directly generalizable to libration by just substituting the overlap $\xi_\mu$ by its corresponding analogue for libration,
\begin{equation}
    \xi_\mu \to \xi_\mu^{(l)}\equiv \int d\Omega_\mathbf{k}\sum_{\pol{\mathbf{k}}} A_{s}(\Omega_\mathbf{k},\pol{\mathbf{k}})A_{\mu}^{(l)}(\Omega_\mathbf{k},\pol{\mathbf{k}}).
\end{equation}

The modification of the recoil heating rates for the libration along the $y$ axis is shown in Fig.~\ref{figure5}(c). The dashed lines, which indicate the case of perfect overlap $\vert\xi_\mu^{(l)}\vert = 1$, are identical to those in Fig.~\ref{Figure2}(a), indicating that the librational recoil heating rate could also be fully suppressed using a suitable angular distribution of the squeezed light $A_s(\Omega_\mathbf{k},\boldsymbol{\varepsilon}_\mathbf{k})$. The colored lines show the recoil heating modification for a squeezed beam propagating along the $z$ axis, polarized along the $y$ axis (see Fig.~\ref{figure5}(a)) and focused onto the particle by a lens of numerical aperture NA. To describe this beam we use the expression in \eqnref{Gaussian} under the substitution $\Theta(-\theta_\mathbf{k}+\pi/2) \to \Theta(\theta_\mathbf{k}-\pi/2)$. At $15$dB and for NA$=0.8$, the librational recoil heating can be suppressed by $40\%$ or increased by a factor $13$, depending on the squeezing phase. These modifications, especially the large increase, could also be measured in current experiments~\cite{vanderLaanPRL2021}. Note that the recoil heating modifications are less pronounced than for the motion along the $z$ axis (compare with Fig.~\ref{Figure2}(a)), due to the lower overlap of the function $A_\mu(\Omega_\mathbf{k},\polK)$ with the squeezed beam. This is evidenced by the differential scattering cross sections in Fig.~\ref{figure5}(d). As shown by this panel, at least half of the photon scattering profile (i.e. across a $\sim 2\pi$ solid angle) remains unaffected by the squeezing. 
A possible way to achieve higher performance in current experiments would be to shine onto the particles two counter-propagating squeezed beams in a standing wave configuration~\cite{MaurerQED2}.

\section{Conclusion}\label{SecConclusion}

To conclude, we have theoretically shown that squeezed light can be used to control the information that the light scattered by an optically levitated nanoparticle carries about its mechanical degrees of freedom. Our formalism applies to both the center-of-mass motion of nanospheres and to librational degrees of freedom of optically trapped nanorotors~\cite{HoangPRL2016,KuhnNatComm2017,AhnPRL2018,AhnNatNano2020,BangPRR2020,DelordNature2020,vanderLaanPRA2020,vanderLaanPRL2021}. Our results predict that one can either reduce measurement back-action noise to boost the coherence time of the quantum mechanical degree of freedom or reduce imprecision noise to perform more efficient displacement measurements in levitated free-space optomechanics. We have shown that this method can be promptly used to reduce displacement measurement sensitivities below the standard quantum limit. Our results are timely as they provide a new tool to improve the quantum control of current experiments cooling optically levitated nanoparticles~\cite{DelicScience2020,MagriniNature2021,TebbenjohannsNature2021,KambaPRA2021,RanfagniPRR2022,KambaOptExp2022}. This in turn paves the way to preparing more complex quantum states (e.g., mechanical squeezed states, non-Gaussian quantum states) which is key to test quantum mechanics and develop quantum-enhanced sensing schemes~\cite{GonzalezBallesteroScience2021,MooreQSciTechnol2021}.

\acknowledgments

We acknowledge discussions with F. Tebbenjohanns. This research was supported by the European Union’s Horizon 2020 research and innovation programme under grant agreement No. [863132] (IQLev) and by the European Research Council (ERC) under the grant Agreement No. [951234] (Q-Xtreme ERC-2020-SyG). This work has been partly supported by ETH Grant No.
ETH-47 20-2.

\appendix

\section{Derivation of motional and librational Hamiltonians}\label{AppA}

In this section of the Appendix we summarize the derivation of the fundamental light-matter Hamiltonian~\eqnref{H0} and its  analogue for nanorotor libration.

\subsection{Center-of-mass Hamiltonian}

We consider a sub-wavelength spherical particle of relative permittivity $\epsilon$, mass $m$, and volume $V$ situated at the origin of coordinates and interacting with the electromagnetic field. The total Hamiltonian $\hat{H}=\hat{\mathbf{p}}^2/(2m) + \hat{H}_{\rm EM} + \hat{H}_{\rm int}$ contains the kinetic energy of the particle with center-of-mass momentum $\hat{\mathbf{p}}$, the Hamiltonian of the electromagnetic field $\hat{H}_{\rm EM}$, and the interaction between the field and the center-of-mass motion $\hat{H}_{\rm int}$. In the small-particle limit the latter two terms can be well approximated in terms of plane-wave modes, labelled by the wave-vector $\mathbf{k}\in\mathds{R}^3$, polarization vectors $\boldsymbol{\varepsilon}_\mathbf{k}\perp\mathbf{k}$, and frequency $\omega (k)=c|\mathbf{k}|$~\cite{MaurerQED1}, as $\hat{H}_{\rm EM} \approx \hbar\int d\mathbf{k}\sum_{\polK} \omega(k)\hat{a}^\dagger(\keps)\hat{a}(\keps)$ and $\hat{H}_{\rm int}=-(\alpha/2)\hat{\mathbf{E}}^2(\hat{\mathbf{r}})$~\cite{RomeroIsartNJP2010,ChangPNAS2010,RomeroIsartPRA2011,GonzalezBallesteroPRA2019} respectively, with $\hat{\mathbf{r}}$ the center-of-mass position operator, $\alpha\equiv3\epsilon_0 V(\epsilon-1)/(\epsilon+2)$ the particle polarizability, and $\hat{\mathbf{E}}(\mathbf{r})$
the electric field operator,
\begin{equation}\label{Eoperator}
    \hat{\mathbf{E}}(\mathbf{r})= i\int d\mathbf{k}\sum_{\polK}\sqrt{\frac{\hbar \omega (k)}{2\epsilon_0 (2\pi)^3}}\left(\boldsymbol{\varepsilon}_\mathbf{k}e^{i\mathbf{k}\mathbf{r}}\hat{a}(\keps)-\text{H.c.}\right).
\end{equation}

Let us now expand the interaction Hamiltonian $\hat{H}_{\rm int}$ to second order in center-of-mass-position operators $\hat{r}_\mu$ with $\mu\in \lbrace x,y,z \rbrace$ (Lamb-Dicke expansion). The zeroth-order term is negligible in the small-particle limit~\cite{MaurerQED1}, whereas the second-order term provides the harmonic trapping potential. The total Hamiltonian is then written as
\begin{equation}
    \hat{H}\approx \hat{H}_p + \hat{H}_{\rm EM}  -\frac{\alpha}{2}(\hat{\mathbf{r}}\cdot\nabla)\hat{\mathbf{E}}^2(\mathbf{0}),
\end{equation}
with $\hat{H}_p=\hbar \sum_\mu\Omega_\mu\hat{b}^\dagger_\mu\hat{b}_\mu$, $\Omega_\mu$ the mechanical frequency, and $\hat{b}_\mu^\dagger$ and $\hat{b}_\mu$ the motional creation and annihilation operators defined through the center-of-mass position operators $\hat{r}_\mu\equiv r_{0\mu}(\hat{b}_\mu+\hat{b}_\mu^\dagger)$ with $r_{0\mu}\equiv[\hbar/(2m\Omega_\mu)]^{1/2}$ the zero-point displacement along axis $\mu$.

We aim at studying the behavior of the system in the presence of (i) an optical  laser, here modelled as a single electromagnetic plane-wave mode $\{\mathbf{k}_0,\pol{\mathbf{k}_0}\}$ whose occupation is very large, and (ii) the squeezed state of a collective mode formed by a linear combination of plane waves. We first account for (i) by applying the following unitary transformation to the Hamiltonian,
\begin{multline}\label{displacementTrafo}
    \hat{T}_1(t) \equiv   \exp\left [\alpha_0^*\hat{a}(\mathbf{k}_0,\pol{\mathbf{k}_0})-\text{H.c.} \right]
    \\ \times\exp\left[i\omega_0t\int d\mathbf{k}\sum_{\polK}\hat{a}^\dagger(\keps)\hat{a}(\keps)\right],
\end{multline}
i.e. a transformation into a frame rotating at the laser frequency $\omega_0$ and a coherent displacement of mode $\{\mathbf{k}_0,\pol{\mathbf{k}_0}\}$ by an amplitude $\alpha_0\in\mathds{C}$. The bosonic electromagnetic operators are transformed as $\hat{T}_1(t)\hat{a}(\keps) T_1^\dagger(t)=e^{-i\omega_0t}[\alpha_0\delta(\mathbf{k}-\mathbf{k}_0)\delta_{\polK\pol{\mathbf{k}_0}}+\hat{a}(\keps)]$. The transformed Hamiltonian can be split into three terms of different orders in $|\alpha_0|$. For $|\alpha_0|\gg1$ we can neglect the $\mathcal{O}(1)$ term thus linearizing the Hamiltonian. Furthermore, assuming the  laser frequency $\omega_0$ is optical, we can neglect rapidly rotating terms at frequencies $\pm 2\omega_0$ under a rotating wave approximation. This results in a linearized Hamiltonian
\begin{multline}\label{LinearizedHamiltonian}
    \hat{H}/\hbar = \int d\mathbf{k}\sum_{\polK}\Delta (k) \hat{a}^\dagger(\keps)\hat{a}(\keps) + \sum_{\mu}\Omega_\mu\hat{b}^\dagger_\mu\hat{b}_\mu \\+\int d\mathbf{k}\sum_{\polK}\hat{q}_\mu\left[G_{\mu}(\keps)\hat{a}^\dagger(\keps)+\text{H.c.}\right],
\end{multline}
with $\hat{q}_\mu\equiv\hat{b}^\dagger_\mu+\hat{b}_\mu$, $\Delta(k) \equiv \omega (k)-\omega_0$, and
$G_{\mu}(\keps)$ the linearized coupling rates. For a laser mode $\{\mathbf{k}_0,\boldsymbol{\varepsilon}_0\}$ propagating along the positive $z-$axis and polarized along the $x-$axis the coupling rates read
\begin{equation}\label{couplingMOTION}
    G_{\mu}(\keps)\equiv i\alpha_0\alpha
    \frac{\sqrt{\omega(k)\omega_{0}}}{2\epsilon_0 (2\pi)^3}
     (\boldsymbol{\varepsilon}_\mathbf{k}^*\cdot \mathbf{e}_x)r_{0\mu}[(\mathbf{k}-k_0\mathbf{e}_z)\cdot\mathbf{e}_\mu].
\end{equation}
The above expression assumes that the gradient of the  laser phase at the trap center is equal to $k_0$, an approximation valid in the Lamb-Dicke regime~\cite{GonzalezBallesteroPRA2019}. At the laser frequency, the coupling rates can be expressed as
\begin{equation}\label{Gidentity}
    G_{\mu}(\keps)\big\vert_{k=k_0} = \sqrt{\frac{c^3\Gamma_\mu^{(0)}}{2\pi \omega_0^2}} A_{\mu}(\Omega_\mathbf{k},\pol{\mathbf{k}}),
\end{equation}
with $\Gamma_\mu^{(0)}$ the recoil heating in the absence of squeezing, $\Omega_{\mathbf{k}}$ the two spherical angles describing the direction of the unit vector $\mathbf{e}_{\mathbf{k}}\equiv\mathbf{k}/k$, and $A_{\mu}(\Omega_\mathbf{k},\pol{\mathbf{k}})$ an orthonormal set of angular distributions ($\mu=x,y,z$) given by
\begin{equation}
    A_{\mu}(\Omega_\mathbf{k},\pol{\mathbf{k}})\equiv i e^{i\arg(\alpha_0)}\sqrt{\frac{3}{8\pi l_\mu}} (\boldsymbol{\varepsilon}_\mathbf{k}\cdot\mathbf{e}_x)\left[(\mathbf{e}_\mathbf{k}-\mathbf{e}_z)\cdot\mathbf{e}_\mu\right],
\end{equation} 
The two remaining undefined quantities, namely the displaced mode amplitude $|\alpha_0|$ and the mechanical frequency $\Omega_\mu$, can be expressed in the paraxial approximation in terms of the laser power $P_t$ and its waist at the focus $W_t$ as~\cite{GonzalezBallesteroPRA2019}
\begin{equation}
    \vert\alpha_0\vert^2 = \frac{16\pi^2P_t}{\hbar c^2k_0 W_t^2},
\end{equation}
\begin{equation}
    \Omega_x=\Omega_y =\Omega_z\frac{\sqrt{2}\pi W_t}{\lambda_0}=
    \sqrt{\frac{\epsilon-1}{\epsilon+2}\frac{12 P_t}{\pi c \rho W_t^4}}
\end{equation}
with $\lambda_0=2\pi k_0^{-1}$ and $\rho=m/V$ the mass density.

\subsection{Hamiltonian for nanorotor libration}

The above derivation can also be applied to the librational degrees of freedom of nanorotors~\cite{HoangPRL2016,KuhnNatComm2017,AhnPRL2018,AhnNatNano2020,BangPRR2020,DelordNature2020,vanderLaanPRA2020,vanderLaanPRL2021}. We consider a sub-wavelength anisotropic particle (e.g. a nanodumbbell) interacting with an optical  laser that is linearly polarized along the $x-$axis and propagating along the positive $z-$axis. The linear electromagnetic response of the particle is modelled by an isotropic relative permittivity $\epsilon$ and a polarizability tensor $\overline{\boldsymbol{\alpha}}$ which, in the body frame, reads $(\overline{\boldsymbol{\alpha}}_{\rm BF})_{ij}=\overline{\alpha}_{ij} = \alpha_i\delta_{ij}$ with $\alpha_1=\alpha_2<\alpha_3$.  We assume that the position of the center of mass of the particle is fixed at the origin, but that the particle is free to rotate about its three axes. The  laser aligns the long axis of the particle with its polarization axis.  We therefore assume that the particle is aligned along the $x-$axis and can undergo small angular displacements $\eta_\mu$ along the axes $\mu\in \lbrace y,z \rbrace$, i.e. small rotations around the $\{z,y\}$ axes. Spinning around the main axis of the particle can be neglected for thin rotors~\cite{SebersonPRA2020}. The Hamiltonian is then given by
\begin{equation}
    \hat{H} = \sum_{\mu}\frac{\hat{L}_\mu^2}{2I} + \hat{H}_{\rm EM} -\frac{1}{2}\hat{\mathbf{E}}(\mathbf{0})\cdot \overline{\boldsymbol{\alpha}}_{\rm LF} \cdot \hat{\mathbf{E}}(\mathbf{0}),
\end{equation}
with $I$ the moment of inertia, assumed equal for rotations along $y$ and $z$ axes, $\hat{L}_{\mu}$ the $\mu-$component of the angular momentum operator, $\hat{H}_{\rm EM}$ the electromagnetic field Hamiltonian as given in \eqnref{HemBARE} in the main text, and $\overline{\boldsymbol{\alpha}}_{\rm LF}$ the polarizability tensor in the laboratory frame. 

We now proceed in similar steps as for the center-of-mass motion. First, we assume the angular displacements are small, and hence expand the polarizability tensor to second order in the angular displacements as $\overline{\boldsymbol{\alpha}}_{\rm LF} = \overline{\boldsymbol{\alpha}}_{\rm BF}+(\hat{\eta}_y,\hat{\eta}_z) \approx (\Delta\alpha)\overline{M}(\hat{\eta}_y,\hat{\eta}_z)$, with $\Delta\alpha\equiv\alpha_3-\alpha_1$ and
\begin{equation}
    \overline{M}(\hat{\eta}_y,\hat{\eta}_z)\equiv  \left[
    \begin{array}{ccc}
        -(\hat{\eta}_z^2+\hat{\eta}_y)^2 & \hat{\eta}_y& \hat{\eta}_z\\
        \hat{\eta}_y&\hat{\eta}_y^2& \hat{\eta}_y\hat{\eta}_z\\
        \hat{\eta}_z&\hat{\eta}_y\hat{\eta}_z&\hat{\eta}_z^2
    \end{array}
    \right].
\end{equation}
The contribution of the constant term $\overline{\boldsymbol{\alpha}}_{\rm BF}$ is again negligible in the small-particle limit which leads to $\overline{\boldsymbol{\alpha}}_{\rm LF}  \approx (\Delta\alpha)\overline{M}(\hat{\eta}_y,\hat{\eta}_z)$.
Using the rotated spherical coordinate system $\{x,y,z\}=r\{\cos\theta,\sin\theta\cos\phi,\sin\theta\sin\phi\}$, and to first order in the polar angle $\theta$, the small displacements can be written as $\eta_y\approx\theta\cos\phi$ and $\eta_z\approx\theta\sin\phi$, allowing to write the angular momenta in the kinetic term as $\hat{L}_z^2+\hat{L}_y^2\approx-\hbar^2[\partial_{\eta_z}^2+\partial_{\eta_y}^2]$. Then, we apply the unitary operator in~\eqnref{displacementTrafo} and linearize the Hamiltonian assuming that the coherent amplitude is large, $\vert\alpha_0\vert\gg1$. Next, we apply the rotating wave approximation and neglect terms oscillating at frequencies $\pm 2\omega_0$. The remaining terms contain (i) a harmonic  potential $(I/2)\Omega^2(\hat{\eta}_z^2+\hat{\eta}_y^2)$ which allows us to define bosonic ladder operators $\hat{b}_\mu$ for the rotational degrees of freedom, via $\hat{\eta}_\mu = r_{0\mu}(\hat{b}_\mu+\hat{b}_\mu^\dagger)$, with the zero-point angular displacement $r_{0\mu}\equiv[\hbar/(2I\Omega)]^{1/2}$ and libration frequency
\begin{equation}\label{eq:librationalFrequency}
    \Omega = \sqrt{\frac{\Delta\alpha}{I}\frac{\hbar\omega_0\vert\alpha_0\vert^2}{\epsilon_0(2\pi)^3}},
\end{equation}
and (ii) a linearized interaction between the librational degrees of freedom and the electromagnetic field. The total Hamiltonian can be written in analogous form to \eqnref{H0}, namely
\begin{multline}\label{Htransformed2}
    \hat{H}^{(0)}=\hbar \sum_\mu\Omega_\mu\hat{b}^\dagger_\mu\hat{b}_\mu + \hat{H}_{\rm EM,r} 
    \\
    +\hbar \int d\mathbf{k}\sum_{\boldsymbol{\varepsilon}_\mathbf{k}}\sum_\mu\hat{q}_\mu\left(G_{\mu}^{(l)}(\keps)\hat{a}^\dagger(\keps) +\text{H.c.}\right),
\end{multline}
with linearized coupling rates for libration (labeled by the super-index $(l)$ to distinguish from the center-of-mass motion analogues)
\begin{equation}\label{couplingROTATION}
    G_{\mu}^{(l)}(\keps)\equiv -\alpha_0 (\Delta\alpha) \frac{\sqrt{\omega(k)\omega_{0}}}{2\epsilon_0 (2\pi)^{3}}r_{0\mu}  (\boldsymbol{\varepsilon}_\mathbf{k}\cdot\mathbf{e}_\mu).
\end{equation} 
In deriving this Hamiltonian we have also neglected a term that reads $\hbar(\hat{\eta}_z^2+\hat{\eta}_y^2)\sum_{\kappa}(g'_{\kappa}\hat{a}^\dagger_\kappa+\text{H.c.})$ as the associated coupling rates are much smaller for typical parameters, namely $\vert g'_{\kappa}\vert/\vert G_{\mu}^{(l)}(\keps)\vert \sim r_{0\mu} \ll 1$.

At the frequency of the  laser light, the coupling rate~\eqnref{couplingROTATION} also takes the form of~\eqnref{GlinearizedGamma0}, i.e.,
\begin{equation}
    G_{\mu}^{(l)}(\keps)\big\vert_{k=k_0} = \sqrt{\frac{c^3\Gamma_\mu^{(0)}}{2\pi \omega_0^2}} A_{\mu}^{(l)}(\Omega_\mathbf{k},\pol{\mathbf{k}}),
\end{equation}
with a recoil heating rate in the absence of squeezing given by
\begin{equation}\label{recoilROTATION}
    \Gamma_\mu^{(0)} = \frac{V}{4\pi^2c}\vert \alpha_0\vert^2\left(\frac{\Delta\alpha}{2\epsilon_0(2\pi)^3}\right)^2\frac{8\pi k_0^2}{3}r_{0\mu}^2\omega_0^2,
\end{equation}
and square-normalized libration angular distributions  given by
\begin{equation}
    A_{\mu}^{(l)}(\Omega_\mathbf{k},\pol{\mathbf{k}})\equiv -e^{i\text{arg}(\alpha_0)}\sqrt{\frac{3}{8\pi}}(\pol{\mathbf{k}}\cdot\mathbf{e}_\mu),
\end{equation}
 which are also orthonormal.

\section{Expressions for input, back-action, and correlation power spectral densities} \label{AppPSD}

In this section of the Appendix we derive the expressions of the power spectral densities used in the main text and state some properties. We first focus on the power spectral densities of the input light, which reflect the modified fluctuations present in the squeezed state. To do so we apply the definition
\begin{equation}
    S_{\mu, AB}(\omega)\equiv  \int \frac{d\tau}{2\pi} e^{i\omega\tau}\langle\hat{A}_\mu(t+\tau)\hat{B}_\mu(t)\rangle,
\end{equation}
to two arbitrary input quadratures $A_\mu,B_\mu\in \{\tilde{X}^{\rm in}_\mu,\tilde{Y}^{\rm in}_\mu\}$. Under the approximations used in this work the spectral densities are straightforward to compute and read
\begin{multline}
    2\pi S_{\mu,XX}^{\rm in}=\frac{\Gamma_\mu}{\Gamma_\mu^{(0)}}\\=1+2\vert\xi_\mu\vert^2 s_0(s_0-c_0\cos(\phi_s-2\psi_\mu)),
\end{multline}
\begin{equation}
    2\pi S_{\mu,YY}^{\rm in}=1+2\vert\xi_\mu\vert^2s_0(s_0+c_0\cos(\phi_s-2\psi_\mu)) ,
\end{equation}
\begin{equation}
    2\pi \left(S_{\mu,XY}^{\rm in}-S_{\mu,YX}^{\rm in}\right)=2i ,
\end{equation}
\begin{equation}
    2\pi \frac{S_{\mu,XY}^{\rm in}+S_{\mu,YX}^{\rm in}}{2}=
    -2 \vert\xi_\mu\vert^2 s_0c_0\sin(\phi_s-2\psi_\mu).
\end{equation}

Let us now focus on the back-action and correlation power spectral densities, defined in Eqs.~\eqref{Sba} and \eqref{Sbacrossed}. By direct computation we find
\begin{multline}\label{SbaSM}
     S_{\mu}^\text{ba}(\omega)\equiv\frac{r_{0\mu}^2}{2\pi}\int_{\mathbb{R}}\text{d}\tau \langle\hat{q}^\text{ba}_{\mu}(t+\tau)\hat{q}^\text{ba}_{\mu}(t)\rangle e^{i \omega\tau}
     \\
     =\frac{r_{0\mu}^2\Gamma_\mu}{2\pi}\left\vert \frac{2\Omega_\mu}{\Omega_\mu^2-\omega^2}\right\vert^2
     ,
\end{multline}
and
\begin{multline}\label{SbacrossedSM}
   S^c_\mu(\omega) \equiv\frac{r_{0\mu}}{2\pi}\int d\tau e^{i\omega\tau}\\ \times \bigg[\langle\tilde{Y}_\mu^{\rm in}(t+\tau)\hat{q}_{\mu}^{\rm ba}(t)\rangle +
   \langle\hat{q}_{\mu}^{\rm ba}(t+\tau)\tilde{Y}_\mu^{\rm in}(t)\rangle\bigg] 
   \\=r_{0\mu}\sqrt{\Gamma_\mu^{(0)}} \frac{2\Omega_\mu}{\Omega_\mu^2-\omega^2}
   \left(S_{\mu,XY}^{\rm in}+S_{\mu,YX}^{\rm in}\right).
\end{multline}
The divergences at $\omega = \Omega_\mu$ are due to the fact that the mechanical damping has been so far neglected. We correct this by noting that, as expected~\cite{bowen2015quantum}, one can write the above expressions as
\begin{equation}
    S_{\mu}^\text{ba}(\omega) =\frac{r_{0\mu}^2\Gamma_\mu}{2\pi}\vert 2m\Omega_{\mu}\chi_{0\mu}(\omega)\vert^2
\end{equation}
and
\begin{equation}
    S^c_\mu(\omega) =r_{0\mu}\sqrt{\Gamma_\mu^{(0)}} \text{Re}[2m\Omega_{\mu}\chi_{0\mu}(\omega)]
   \left(S_{\mu,XY}^{\rm in}+S_{\mu,YX}^{\rm in}\right)
\end{equation}
in terms of the zero-loss susceptibility
\begin{equation}
	\chi_{0\mu}(\omega)\equiv\lim_{\gamma_\mu\to 0^+}\chi_{\mu}(\omega) \equiv \lim_{\gamma_\mu\to 0^+}\frac{1}{m}\frac{1}{\Omega_{\mu}^2-\omega^2-i\gamma_\mu\omega},
\end{equation}
with $\chi_{\mu}(\omega)$ the mechanical susceptibility. This allows us to extend the above results to the case of non-zero damping by substituting by the real susceptibility, i.e. $\chi_{0\mu}(\omega) \to \chi_{\mu}(\omega)$.

\section{Wigner function of input light}\label{AppWigner}

In this final section of the Appendix we derive the Wigner function of the input light shown in the insets of \figref{Figure1}(b) and \figref{figure4}(a). To do so we define detection modes in terms of the interacting modes as~\cite{AndreiArxiv2022}
\begin{equation}
    \hat{a}_\mu \equiv \int_{-\infty}^{\infty}dt h(t)\tilde{a}_\mu (t),
\end{equation}
with a square-normalized function $h(t)$ guaranteeing bosonic commutation relations $[\hat{a}_\mu,\hat{a}_{\mu}^\dagger]=1$ for both input and output detection modes. Without loss of generality we choose for $h(t)$ a modulated step function,
\begin{equation}
    h(t) = \Theta(-t+T/2)\Theta(t+T/2)e^{i\tilde{\Omega} t}/\sqrt{T},
\end{equation}
with $\tilde{\Omega}$ an arbitrary frequency and $1/T$ a detection bandwidth which, in the spirit of the slowly varying couplings and squeezing functions, is taken much wider than the chosen frequency, $\tilde{\Omega} T \ll 1$. We can now define amplitude and phase quadratures $\hat{X}_{\mu} = \hat{a}^\dagger_{\mu}+\text{H.c.}$, $\hat{Y}_{\mu} = i\hat{a}_{\mu}^{\dagger}+\text{H.c.}$ and, using the commutation relation of the interacting modes~\cite{ParisPRA2003}, derive the Wigner function
\begin{equation}\label{Wigner}
    W_\mu^{\rm in}(X,Y) = \frac{1}{2\pi\sqrt{\text{det}[\overline{C}_\mu]}}\exp\left[-\frac{\mathbf{X}_\mu\cdot \overline{C}_\mu^{-1}\cdot\mathbf{X}_\mu}{2 }\right],
\end{equation}
with $\mathbf{X}_\mu\equiv [X_\mu,Y_\mu]^T$ and $\overline{C}_\mu$ the unnormalized covariance matrix,
\begin{equation}
    \overline{C}\equiv\left[
    \begin{array}{cc}
        2\pi S_{\mu,XX}^{\rm in} & -2\pi S_{\rm cross}^{\rm in}  \\
        -2\pi S_{\rm cross}^{\rm in} & 2\pi S_{\mu,YY}^{\rm in}
    \end{array}
    \right],
\end{equation}
where $S_{\mu, \rm cross}^{\rm in} \equiv (S_{\mu,XY}^{\rm in}+S_{\mu,YX}^{\rm in})/2.$ The Wigner function~\eqnref{Wigner} is displayed in the inset of \figref{figure4}(a) in the main text. A similar expression can be derived using the bare squeezed detection mode
\begin{equation}
    \hat{a}(\omega) \equiv \frac{1}{\sqrt{\epsilon}}\int_{\omega-\epsilon/2}^{\omega+\epsilon/2} d\omega'\hat{a}_s(\omega'),
\end{equation}
with $\epsilon$ small enough such that $r(\omega \pm\epsilon/2)\approx r(\omega)$ and $\phi(\omega\pm\epsilon/2)\approx\phi(\omega)$. We define quadratures $\hat{Q}(\omega) \equiv \hat{a}^\dagger(\omega) + \text{H.c.}$ and $\hat{P}(\omega) \equiv i[\hat{a}^\dagger(\omega)-\text{H.c.}]$, and compute the following expected values in the squeezed vacuum state,
\begin{equation}
    \langle \hat{Q}^2(\omega)\rangle = e^{2r(\omega)}\cos^2[\phi(\omega)/2]+e^{-2r(\omega)}\sin^2[\phi(\omega)/2],
\end{equation}
\begin{equation}
    \langle \hat{P}^2(\omega)\rangle = e^{2r(\omega)}\sin^2[\phi(\omega)/2]+e^{-2r(\omega)}\cos^2[\phi(\omega)/2],
\end{equation}
\begin{equation}
    \langle \hat{Q}(\omega)\hat{P}(\omega)+\hat{P}(\omega)\hat{Q}(\omega)\rangle = -2\sinh[2r(\omega)]\sin[\phi(\omega)].
\end{equation}
Using these equations we can construct the Wigner function of the electromagnetic field at frequency $\omega$ in full analogy with~\eqnref{Wigner}. This function is shown in \figref{Figure1}(b) in the main text.

\bibliography{main}



\end{document}